\documentclass[twocolumn,english,superscriptaddress,showpacs]{revtex4-2}
\usepackage[T1]{fontenc}
\usepackage[utf8]{inputenc}
\usepackage{babel}
\usepackage{array}
\usepackage{verbatim}
\usepackage{booktabs}
\usepackage{bm}
\usepackage{multirow}
\usepackage{amsmath}
\usepackage{amssymb}
\usepackage{graphicx}
\usepackage[unicode=true,pdfusetitle,
 bookmarks=true,bookmarksnumbered=true,bookmarksopen=true,bookmarksopenlevel=1,
 breaklinks=false,pdfborder={0 0 0},pdfborderstyle={},backref=false,colorlinks=false]
 {hyperref}

\makeatletter

\providecommand{\tabularnewline}{\\}

\usepackage{braket}
\usepackage{bm}
\usepackage{tikz}
\usepackage{pgfplots}
\usepackage{pgf}
\usepackage{subfigure}

\makeatother

\begin{document}
\title{Emergence of Dirac Composite Fermions: Dipole Picture}
\author{Guangyue Ji}
\affiliation{International Center for Quantum Materials, Peking University, Beijing
100871, China}
\author{Junren Shi}
\email{junrenshi@pku.edu.cn}

\affiliation{International Center for Quantum Materials, Peking University, Beijing
100871, China}
\affiliation{Collaborative Innovation Center of Quantum Matter, Beijing 100871,
China}
\begin{abstract}
Composite fermions (CFs) are the particles underlying the novel phenomena
observed in partially filled Landau levels. Both microscopic wave
functions and semi-classical dynamics suggest that a CF is a dipole
consisting of an electron and a double $2h/e$ quantum vortex, and
its motion is subject to a Berry curvature that is uniformly distributed
in the momentum space. Based on the picture, we study the electromagnetic
response of composite fermions. We find that the response in the long-wavelength
limit has a form identical to that of the Dirac CF theory. To obtain
the result, we show that the Berry curvature contributes a half-quantized
Hall conductance which, notably, is independent of the filling factor
of a Landau level and not altered by the presence of impurities. The
latter is because CFs undergo no side-jumps when scattered by quenched
impurities in a Landau-level with the particle-hole symmetry. The
remainder of the response is from an effective system that has the
same Fermi wavevector, effective density, Berry phase, and therefore
long-wavelength response to electromagnetic fields as a Dirac CF system.
By interpreting the half-quantized Hall conductance as a contribution
from a redefined vacuum, we can explicitly show the emergence of a
Dirac CF effective description from the dipole picture. We further
determine corrections due to electric quadrupoles and magnetic moments
of CFs and show deviations from the Dirac CF theory when moving away
from the long wavelength limit.
\end{abstract}
\maketitle

\section{Introduction\label{sec:Introduction}}

The fractional quantum Hall (FQH) effect, a phenomenon discovered
nearly four decades ago~\citep{tsui_two-dimensional_1982}, remains
unique as the only topological effect driven by electron correlations
and observed in laboratories. In two dimensions, electrons subjected
to a strong magnetic field are confined in a Landau level (LL). In
such a system, the kinetic energy is completely quenched, and the
electron-electron interaction dominates its behavior. Surprisingly,
according to the theory of composite fermions~\citep{jain2007composite},
the system can be interpreted as a non-interacting or weakly interacting
system consisting of fictitious particles called composite fermions
(CFs). It is hypothesized that these particles are bound states of
electrons and quantum vortices, reside in a hidden Hilbert space,
and form simple quantum states~\citep{jain_beyond_2009}. The theory
of CFs prescribes an ansatz for mapping a state in the hidden Hilbert
space to a quantum state of a partially filled LL in the real world.
The theory is very successful: by interpreting FQH states as integer
quantum Hall states of free CFs, wave functions constructed from the
ansatz reliably achieve high overlaps with those determined by exact
diagonalizations. The theory also prompts experimental searches for
CFs. In various geometric resonance experiments and ballistic transport
experiments~\citep{willet1998,smet1998,willett_geometric_1999,smet_commensurate_1999},
CFs do exhibit behaviors expected for genuine particles. One could
treat CFs as if they are elementary particles in the hidden Hilbert
space, just like electrons are to the real world.

To treat CFs as the elementary particles of the hidden Hilbert space,
one needs an effective theory to specify their dynamics. For a long
time, the de facto standard effective theory for CFs is the one proposed
by Halperin, Lee, and Read (HLR)~~\citep{halperin_theory_1993}.
The theory underlies the designs and interpretations of most CF experiments~\citep{willet1998,smet1998}.
In the HLR theory, CFs feel an effective magnetic field and behave
just like ordinary particles by following the ordinary Newtonian dynamics.
Such a picture naturally emerges from the Chern-Simons (CS) field
theory of CFs which tries to justify the CF picture by interpreting
the electron-vortex binding as a singular CS gauge transformation~\citep{zhang_effective-field-theory_1989,lopez_fractional_1991,kalmeyer_metallic_1992}.
However, the CS theory disregards the fact that the physics occurs
essentially in the projected Hilbert space of a partially filled LL.
It contains artifacts which have to be eliminated in the HLR theory.
For instance, to remove the unwelcome presence of the bare band mass
of electrons without introducing inconsistencies, one has to assume
phenomenologically that a mass renormalization is accompanied by a
corresponding renormalization to the effective interaction between
CFs~\citep{simon_finite-wave-vector_1993,simon1998}. Another obvious
flaw, i.e., the absence of a half-quantized CF Hall conductance expected
for a half-filled LL with the particle-hole symmetry~\citep{kivelson_composite-fermion_1997},
is remedied only recently by Wang et al., who show that spatial fluctuations
of the effective magnetic field can induce CF scattering with side
jumps and gives rise to a half-quantized CF Hall conductance~\citep{wang_particle-hole_2017}.

These difficulties prompt Son to propose the Dirac CF theory~\citep{son_is_2015}.
Besides the aforementioned issues, Son also notices that the HLR theory
does not exhibit explicitly the particle-hole symmetry expected for
a LL in the zero-mass limit. As an effort to address the issue, Son
conjectures that CFs are massless Dirac particles. The interpretation
of CFs in the Dirac CF theory differs in many ways from that in the
HLR theory. Most notably, a Dirac CF traversing a Fermi circle acquires
a $\pi$-Berry phase, which is absent in the HLR theory. It leads
to the predictions of suppression of back-scattering of CFs~\citep{geraedts_half-filled_2016}
and an extra phase in SdH oscillations~\citep{pan_berry_2017}. Moreover,
the theory shows the swap of the notions of density and magnetic field,
i.e., the density of Dirac CFs is set by the magnetic field, while
the effective magnetic field felt by CFs is set by the electron density.
This is because Dirac CFs are actually particles dual to electrons
in a charge-neutral massless Dirac cone~\citep{mross2016}. This
is distinct from the HLR theory in which CFs remain electron-like,
having the same density, Fermi wavevector and dynamics.

One of manifest distinctions between the two effective theories is
in how the conductivity tensor of electrons $\sigma(\omega)$ is related
to that of CFs $\tilde{\sigma}(\omega)$. In the HLR theory, the two
conductivity tensors are related by a simple resistivity-shifting
rule~\citep{halperin_theory_1993,jain2007composite}
\begin{equation}
\sigma^{-1}(\omega)=\tilde{\sigma}^{-1}(\omega)+\sigma_{\mathrm{CS}}^{-1},\label{eq:HLRrhoTrans}
\end{equation}
with
\begin{equation}
\sigma_{\textrm{CS}}=\frac{e^{2}}{2h}\left(\begin{array}{cc}
0 & 1\\
-1 & 0
\end{array}\right),
\end{equation}
whereas in the Dirac CF theory, they are related by~\citep{son_is_2015}
\begin{equation}
\sigma(\omega)=-\sigma_{\textrm{CS}}\tilde{\sigma}^{-1}(\omega)\sigma_{\textrm{CS}}+\sigma_{\textrm{CS}}\beta(\omega)\label{eq:DiracSigmaTrans}
\end{equation}
with
\begin{equation}
\beta(\omega)=\left(\begin{array}{cc}
1 & -\mathrm{i}\omega/\tilde{\omega}_{B}\\
\mathrm{i}\omega/\tilde{\omega}_{B} & 1
\end{array}\right),\label{eq:beta}
\end{equation}
where $\tilde{\omega}_{B}\equiv eB/m^{\ast}$ is a characteristic
frequency set by the external magnetic field $\bm{B}=-B\hat{z}$.
The relation is established in the long wavelength limit, and depends
on a CF effective mass parameter $m^{\ast}$ which is phenomenologically
introduced to recover the Galilean invariance for the Dirac CF theory~\citep{son_is_2015,levin2017}.
As a result of the distinction, the two theories will predict different
electromagnetic (EM) responses of CF systems.

An alternative effective description parallel to these theories is
the dipole picture proposed by Read~\citep{read_theory_1994}. In
the picture, a CF is not a point particle but a dipole consisting
of an electron and a double $2h/e$ quantum vortex which is spatially
separated from the electron. Intriguingly, the momentum of a CF, which
is proportional to its velocity, can be interpreted as the transverse
of its dipole vector. The picture emerges naturally from the Rezayi-Read
wave function $\Psi(\{z_{a}\})\propto J(\{z_{a}+\mathrm{i}k_{a}l_{\mathrm{B}}^{2}\})\exp(\mathrm{i}\sum_{a}\bar{k}_{a}z_{a}/2)$
which is obtained by applying the CF ansatz $\Psi(\{z_{a}\}=\hat{P}_{\mathrm{LLL}}J(\{z_{a}\})\psi_{\mathrm{CF}}(\{\bm{r}_{a}\})$
to a CF wave function $\psi_{\mathrm{CF}}(\{\bm{r}_{a}\})=\exp(\mathrm{i}\sum_{a}\bm{k}_{a}\cdot\bm{r}_{a})$
for a set of free CFs with wavevectors $\{\bm{k}_{a}\}$ in the hidden
Hilbert space~\citep{rezayi_fermi-liquid-like_1994}, where we denote
$z_{a}=x_{a}+\mathrm{i}y_{a}$ for $\bm{r}_{a}\equiv(x_{a},y_{a})$,
$k_{a}=k_{ax}+\mathrm{i}k_{ay}$ for $\bm{k}_{a}\equiv(k_{ax},k_{ay})$
and $\bar{k}_{a}$ the complex conjugate of $k_{a}$, $J(\{z_{a}\})\propto\prod_{a<b}(z_{a}-z_{b})^{2}$
is the Bijl-Jastrow factor, $l_{B}=\sqrt{\hbar/eB}$ is the magnetic
length, and $\hat{P}_{\mathrm{LLL}}$ is the projection to the lowest
LL~\citep{jain2007composite,jain_beyond_2009}. By interpreting zeros
in the Bijl-Jastrow factor as quantum vortices, one observes that
a double-vortex is displaced from the electron to which it is bound
by $z_{a}\rightarrow z_{a}+\mathrm{i}k_{a}l_{B}^{2}$. Microscopic
wave functions prescribed by the theory of CFs are well tested and
widely accepted as precise descriptions of many-body states of fractionally
filled LLs~\citep{balram_nature_2016}. The fact that the dipole
picture is directly inferred from a microscopic wave function distinguishes
it from the HLR theory and the Dirac CF theory, both of which are
based on conjectured effective field theories and cannot be directly
associated with actual microscopic wave functions except for a few
special cases~\citep{shankar_towards_1997,shankar_hamiltonian_1999,murthy_hamiltonian_2003,gocanin_microscopic_2021}.

For a long time, the dynamics of CFs as dipoles is not explicitly
specified. There is even a misunderstanding that the dipole picture
is just a complement to the HLR theory and has the same dynamics.
To this end, Shi\textit{ }and Ji\textit{ }derive the semi-classical
dynamics of CFs in a CF Wigner crystal that is basically a set of
CF wave packets in the hidden Hilbert space~\citep{shi_dynamics_2018}.
From the semi-classical dynamics, they obtain a dipole picture similar
to that suggested by Read. Specifically, in the picture, CS-like fields
emerge as a result of the correlations embodied in the Bijl-Jastrow
factor and couple only to vortices, while external EM fields couple
only to electrons. By adopting the same interpretation of the momentum
as in Read's theory, they obtain a dynamics in which CFs are subject
to a uniform Berry curvature $\Omega_{z}=1/eB$ in the momentum space~\citep{sundaram_wave-packet_1999,xiao_berry_2010}.
The presence of the Berry curvature is a clear indication that a dipole
is not identical to the CF envisioned in the HLR theory.

Notably, even though the dipole picture and the Dirac CF theory are
very different microscopically, their CFs share some basic properties
(see Sec.~\ref{subsec:EOMs}): (1) a CF traversing a Fermi circle
acquires a $\pi$-Berry phase; (2) the Fermi wavevector $k_{\mathrm{F}}=\sqrt{eB/\hbar}$
is set only by the magnetic field and independent of the electron
density. These properties underly various effects predicted for the
Dirac CF theory~\citep{geraedts_half-filled_2016,pan_berry_2017,ji_asymmetry_2020}.
On the other hand, different from the Dirac CF theory, the dipole
picture does not exhibit explicitly the particle-hole symmetry. It
is due to the lack of the particle-hole symmetry of the vacuum assumed
in the picture: the vacuum is an empty LL for electrons and a filled
LL for holes. The latter has a nonzero Chern number $C_{\mathrm{vac}}=1$.
The difference is superficial and may not have a physical consequence
(see Sec.~\ref{subsec:Emergence-of-the}). Therefore, it is reasonable
to believe that although the two pictures are distinct microscopically,
they may share predictions for macroscopic effects, e.g., the EM response.

In this paper, we determine the EM response of CFs in the dipole picture.
We show that, in the long-wavelength limit, the response has a form
identical to Eq.~(\ref{eq:DiracSigmaTrans}) of the Dirac CF theory,
and is different from Eq.~(\ref{eq:HLRrhoTrans}) even though the
dipole picture is closer microscopically to the HLR theory. The Dirac
CF-like EM response arises because the Berry curvature contributes
a half-quantized intrinsic Hall conductance which, notably, is independent
of the filling factor. We show that CFs undergo no side jumps when
scattered in a LL with the particle-hole symmetry. Therefore, the
intrinsic contribution due to the Berry curvature is not altered by
extrinsic impurity scattering. The remainder of the response is from
an effective system with a Fermi wavevector $k_{\mathrm{F}}=\sqrt{eB/\hbar}$
and a $\pi$-Berry phase, indistinguishable from a Dirac CF system
in the long-wavelength limit. By redefining the vacuum of the dipole
picture and interpreting the intrinsic response as a contribution
from the vacuum, we can explicitly show the emergence of a Dirac CF
effective description. On the other hand, although we can reproduce
the result by Wang et~al. at half filling~\citep{wang_particle-hole_2017},
the dipole theory and the HLR theory are in general not equivalent
when the filling factor of a LL deviates from $1/2$. Finally, we
show that deviations from the Dirac CF theory arise when higher order
corrections due to electric quadrupoles and magnetic moments of CFs
are considered.

The remainder of the paper is organized as follows. In Sec.~\ref{sec:Dipole-picture},
we introduce the dipole picture of CFs, and discuss its dynamics as
well as a constraint imposed by the particle-hole symmetry. In Sec.~\ref{sec:Intrinsic-responses},
we determine the intrinsic response of CFs, and show that it contributes
a half quantized Hall conductance independent of the filling factor.
In Sec.~\ref{sec:Scattering of CF}, we show that CFs in a particle-hole
symmetric LL undergo no side-jumps when scattered. Therefore, the
intrinsic response is not altered by the scattering. In Sec.~\ref{sec:extrinsic responses},
we solve the Boltzmann equation and determine the extrinsic response
of CFs. In Sec.~\ref{sec:Electromagnetic-responses}, we apply CS
self-consistent conditions and determine the total response of CFs
to EM fields, and show the emergence of the Dirac CF effective theory
in the dipole picture. In Sec.~\ref{sec:Quadrupole-and-magnetic},
we determine corrections due to electric quadrupoles and magnetic
moments of CFs. In Sec.~\ref{sec:An-experimental-test}, we test
our theory by fitting to experimental data. Finally, we summarize
and discuss our results in Sec.~\ref{sec:concluding-remarks}. 

\section{Dipole picture of CFs\label{sec:Dipole-picture}}

\subsection{Lagrangian\label{subsec:Lagrangian}}

In the dipole picture, a CF is the bound state of an electron and
a double quantum vortex. The electron and the vortex are separated
spatially, and bounded together by an attractive force due to the
void created by the vortex and the repulsive Coulomb interaction between
electrons~\citep{read_theory_1994}. CS-like fields emerge and couple
to vortices. Meanwhile, electrons are coupled to external EM fields
including the strong magnetic field which confines electrons in a
LL. This is the dipole picture explicated in Ref.~\citep{shi_dynamics_2018}.
Our study will be based on this particular dipole picture instead
of other variants which can be found in literatures~\citep{lee_neutral_1998,lee_chern-simons_1999,pasquier_dipole_1998,von_oppen_half-filled_1999,wang_half-filled_2016}.

The picture of CFs is described by the action in (2+1) dimensions
\begin{equation}
S=\int\mathrm{d}^{3}x\left(-A\cdot j^{\mathrm{e}}-a\cdot j+\frac{e^{2}}{4h}\epsilon^{\mu\nu\gamma}a_{\mu}\partial_{\nu}a_{\gamma}-\varepsilon\right),\label{eq:dipoleaction}
\end{equation}
where $\left(j^{\mathrm{e}}\right)^{\mu}\equiv(\rho^{\mathrm{e}},\bm{j}^{\mathrm{e}})$
and $j^{\mu}\equiv(\rho,\bm{j})$ are charge-current densities of
electrons and vortices, $A^{\mu}\equiv(\Phi,\bm{A})$ and $a^{\mu}\equiv(\phi,\bm{a})$
denote the external EM fields and the emergent CS fields, respectively,
and $\varepsilon$ denotes the binding energy between electrons and
quantum vortices. The action assumes that both electrons and vortices
have zero mass and are subject to the external magnetic field and
the CS magnetic field, respectively. As a result, the motion of the
electrons (vortices) is confined in a LL created by the external (CS)
magnetic field.

Under the mean-field approximation, differentiating the action with
respect to the CS fields $a$ gives rise to the self-consistent conditions
\begin{align}
\rho & =-\frac{e^{2}}{2h}b,\label{eq:btorho}\\
\bm{j} & =-\sigma_{\mathrm{cs}}\bm{e},\label{eq:jtoe}
\end{align}
where $\bm{e}=-\partial_{t}\bm{a}-\bm{\nabla}\phi$ and $\bm{b}=\bm{\nabla}\times\bm{a}\equiv b\hat{z}$
are CS electric and magnetic fields, respectively. While the HLR theory
has the same set of self-consistent conditions, the dipole picture
differs in that (a) the CS fields only couple to vortices and are
not equivalent in effects to the EM fields; (b) the conditions as
well as the CS term in Eq.~(\ref{eq:dipoleaction}) should be regarded
as an approximation only for the long-wavelength limit. Corrections
are expected when the wavelength of the EM fields is comparable to
the length scale of the dipoles, i.e., the magnetic length $l_{\mathrm{B}}$~\citep{shi_dynamics_2018}.

To proceed, we relate the charge-current densities of electrons $j^{\mathrm{e}}$
to their counterparts for vortices $j$. The charge-current densities
of electrons are defined by
\begin{align}
\rho^{\mathrm{e}}\left(t,\bm{x}\right) & =-e\sum_{a}\delta\left(\bm{x}-\bm{x}_{a}^{\mathrm{e}}(t)\right),\\
\bm{j}^{\mathrm{e}}\left(t,\bm{x}\right) & =-e\sum_{a}\dot{\bm{x}}_{a}^{\mathrm{e}}\delta\left(\bm{x}-\bm{x}_{a}^{\mathrm{e}}(t)\right),
\end{align}
where $\{\bm{x}_{a}^{\mathrm{e}}(t)\}$ denotes the set of electron
coordinates. The charge-current densities for vortices $(\rho,\bm{j})$
can be similarly defined by using the set of vortex coordinates $\{\bm{x}_{a}(t)\}$.
We introduce the dipole vector
\begin{equation}
\bm{d}_{a}\equiv\bm{x}_{a}^{\textrm{e}}-\bm{x}_{a}.\label{eq:dipole vector}
\end{equation}
 After applying multipole expansions~\citep{jackson1999}, we obtain
\begin{align}
\rho^{\mathrm{e}} & =\rho-\bm{\nabla}\cdot\bm{P}+\bm{\nabla}\bm{\nabla}:\mathbb{Q}\dots\,,\label{eq:je2-1-1}\\
\bm{j}^{\textrm{e}} & =\bm{j}+\partial_{t}\left(\bm{P}-\bm{\nabla}\cdot\mathbb{Q}\right)+\bm{\nabla}\times\bm{M}+\dots\,,\label{eq:jeexpansion}
\end{align}
where we keep dipole and quadrupole corrections for the charge density,
and displacement current and magnetic dipole corrections for the current
density. Because CFs have a length scale $l_{B}$ and an energy scale
$\hbar\tilde{\omega}_{B}$, the expansions are in rising orders of
$ql_{B}$ and $\omega/\tilde{\omega}_{B}$, where $\omega$ and $q$
are the frequency and wavenumber of probing EM fields, respectively.
The dipole density $\bm{P}$, the quadrupole density tensor $\mathbb{Q}$,
and the magnetization density $\bm{M}$ are defined, respectively,
by
\begin{align}
\bm{P}\left(t,\bm{x}\right) & =-e\sum_{a}\bm{d}_{a}\delta\left(\bm{x}-\bm{x}_{a}\right),\label{eq:polarization-1-1}\\
\mathbb{Q}\left(t,\bm{x}\right) & =-\frac{e}{2}\sum_{n}\bm{d}_{a}\bm{d}_{a}\delta(\bm{x}-\bm{x}_{a}),\\
\bm{M}\left(t,\bm{x}\right) & =-e\sum_{a}\bm{d}_{a}\times\left(\dot{\bm{x}}_{a}+\frac{1}{2}\dot{\bm{d}}_{a}\right)\delta\left(\bm{x}-\bm{x}_{a}\right).\label{eq:magnetization-1-1}
\end{align}
Substituting the expansions into Eq.~(\ref{eq:dipoleaction}) and
applying integrals by parts, we obtain
\begin{multline}
S\approx\int\mathrm{d}^{2}\bm{x}\mathrm{d}t\left[-\left(a+A\right)\cdot j+\frac{e^{2}}{4h}\epsilon^{\mu\nu\gamma}a_{\mu}\partial_{\nu}a_{\gamma}\right.\\
\left.+\bm{M}\cdot\bm{B}-\left(\varepsilon-\bm{E}\cdot\bm{P}-\mathbb{Q}:\bm{\nabla}\bm{E}\right)\right].\label{eq:Sfirstorder}
\end{multline}

We then follow Read to define the momentum of a CF as~\citep{read_theory_1994}
\begin{equation}
\bm{p}_{a}=eB\hat{z}\times\bm{d}_{a}.\label{eq:momentum}
\end{equation}
We find
\begin{equation}
\bm{M}\cdot\bm{B}=\sum_{a}\bm{p}_{a}\cdot\dot{\bm{x}}_{a}+\frac{1}{2eB}\left(\bm{p}_{a}\times\dot{\bm{p}}_{a}\right)\cdot\hat{z}.
\end{equation}
Substituting it into Eq.~(\ref{eq:Sfirstorder}) and interpreting
$\bm{x}_{a}$ as the coordinate of a CF, we obtain an action in which
a CF is governed by the Lagrangian
\begin{multline}
L_{a}=\bm{p}_{a}\cdot\dot{\bm{x}}_{a}+\frac{1}{2eB}\left(\bm{p}_{a}\times\dot{\bm{p}}_{a}\right)\cdot\hat{z}-e\dot{\bm{x}}_{a}\cdot\tilde{\bm{a}}(t,\bm{x}_{a})\\
+e\tilde{\phi}(t,\bm{x}_{a})-\mathcal{E}_{a}.\label{eq:lagrangian}
\end{multline}
From the Lagrangian, we see that the CF is subject to an effective
field $\tilde{a}\equiv(\tilde{\phi},\tilde{\bm{a}})=a+A$, a Berry
curvature $\Omega_{z}=1/eB$, and an energy dispersion with electric
dipole and quadrupole corrections
\begin{multline}
\mathcal{E}_{a}=\varepsilon\left(\bm{p}_{a}\right)-\frac{1}{B}\left(\bm{E}\times\hat{z}\right)\cdot\bm{p}_{a}\\
+\frac{1}{2eB^{2}}\left(\bm{p}_{a}\times\hat{z}\right)\cdot\bm{\nabla}\bm{E}\cdot\left(\bm{p}_{a}\times\hat{z}\right),\label{eq:Epsilon}
\end{multline}
where $\varepsilon\left(\bm{p}_{a}\right)$ denotes the binding energy
of a CF, and $\varepsilon=\sum_{a}\varepsilon\left(\bm{p}_{a}\right)$.

We model the binding energy as a harmonic potential $\varepsilon\left(\bm{p}_{a}\right)\propto|\bm{d}_{a}|^{2}$.
After substituting Eq.~(\ref{eq:momentum}), the binding energy becomes
the dispersion of CFs, and can be written as
\begin{equation}
\varepsilon\left(\bm{p}_{a}\right)=\frac{b}{B}\frac{p_{a}^{2}}{2m^{*}},\label{eq:dispersion}
\end{equation}
where we interpret $m^{*}$ as the effective mass of CFs. For a reason
which will be clarified in Sec.~\ref{subsec:Particle-hole symmetry},
we append a $b/B$ factor to the dispersion. %
The extra factor becomes 1 at half filling.

\subsection{Dynamics\label{subsec:EOMs}}

From the Lagrangian Eq.~(\ref{eq:lagrangian}), we can obtain equations
of motion of CFs. For the moment, we focus on the long-wavelength
limit and ignore all spatial gradients of $\bm{E}$, $\bm{B}$ and
$\bm{b}$. The time dependence of $\bm{B}$ is also ignored since
$\partial_{t}\bm{B}=-\bm{\nabla}\times\bm{E}$ is of the same order
of the gradient of $\bm{E}$. The equations read:
\begin{align}
\dot{\bm{x}} & =\frac{b}{B}\frac{\bm{p}}{m^{\ast}}-\frac{1}{B}\bm{E}\times\hat{z}-\frac{1}{eB}\dot{\bm{p}}\times\hat{z},\label{eq:EOMxdot}\\
\dot{\bm{p}} & =-e\tilde{\bm{e}}-e\tilde{b}\dot{\bm{x}}\times\hat{z},\label{eq:EOMpdot}
\end{align}
where for the simplicity of notations we drop the subscripts indexing
particles, $\tilde{\bm{e}}=\bm{e}+\bm{E}$ and $\tilde{b}=b-B$ are
the effective electric and magnetic fields experienced by CFs, respectively.
The equations should be regarded as a zeroth order approximation.
Corrections due to gradients of the EM and CS fields will be determined
in Sec.~\ref{sec:Quadrupole-and-magnetic}.

The dynamics of CFs in the dipole picture is different from that assumed
in the HLR theory. Two differences are notable: (a) CFs are subject
to a Berry curvature $\Omega_{z}=1/eB$; (b) The external electric
field $\bm{E}$ introduces a correction to the group velocity due
to the dipole correction to the energy. In the HLR theory, EM fields
can be completely absorbed into the effective fields. Therefore, they
are equivalent to the CS fields in driving CFs. This is not true anymore
in the dipole picture. Besides being a part of the effective fields,
they also introduce the Berry curvature and the dipole correction.
As we will see, these corrections change fundamentally how a CF system
responses to EM fields.

The presence of the Berry curvature has important physical consequences.
It introduces a $\pi$-Berry phase for an electron traversing a Fermi
circle. The $\pi$-Berry phase is regarded as a feature of the Dirac
CF theory, and its effects has been extensively explored~\citep{geraedts_half-filled_2016,pan_berry_2017}.
Another consequence of the Berry curvature is a correction to the
phase-space density-of-states~\citep{xiao_berry_2005}. In ordinary
two-dimensional systems, a quantum state always occupies a $h^{2}$
phase-space volume. In the presences of both the Berry curvature and
the (effective) magnetic field, however, the phase space volume element
$\mathrm{d}^{2}x\mathrm{d}^{2}p/h^{2}$ is modified to $D\mathrm{d}^{2}x\mathrm{d}^{2}p/h^{2}$
by the density-of-state correction factor
\begin{equation}
D=1+e\tilde{b}\Omega_{z}=\frac{b}{B}.
\end{equation}
For a homogeneous system, we have $D=2\nu$, where $\nu\equiv-\rho^{\mathrm{e}}h/e^{2}B$
is the filling factor of the system. The factor modifies the Fermi
wavevector from its usual value $k_{\mathrm{F}}^{0}=\sqrt{4\pi n}$,
where $n$ is the density of CFs, to
\begin{equation}
k_{\mathrm{F}}=k_{\mathrm{F}}^{0}/\sqrt{D}=\sqrt{\frac{eB}{\hbar}}.
\end{equation}
We see that, even though the CF density in the dipole picture is the
same as the electron density, the Fermi wavevector is set only by
the magnetic field $B$ and independent of the density. The prediction
differs from the HLR theory and agrees with the Dirac CF theory but
has a different interpretation: $k_{\mathrm{F}}$ is modified due
not to a change of the CF density but the phase-space density-of-state
correction.

For the convenience of later applications, the equations of motion
Eqs.~(\ref{eq:EOMxdot}, \ref{eq:EOMpdot}) can be solved for $\dot{\bm{x}}$
and $\dot{\bm{p}}$:
\begin{align}
\dot{\bm{x}}= & \frac{\bm{p}}{m^{*}}-\frac{1}{b}\hat{z}\times\bm{e},\label{eq:EOM3_1}\\
\dot{\bm{p}}= & e\tilde{b}\hat{z}\times\frac{\bm{p}}{m^{*}}-\frac{1}{D}e\bm{e}-e\bm{E}.\label{eq:EOM3_2}
\end{align}

\subsection{Particle-hole symmetry and CF dispersion\label{subsec:Particle-hole symmetry}}

The particle-hole symmetry imposes a constraint on how the CF dispersion
depends on the filling factor. To see that, we assume that CFs have
the dispersion $\varepsilon_{\bm{p}}=\bm{p}^{2}/2m^{\prime}$ as in
the HLR theory. Away from half-filling, CFs experience an effective
magnetic field $\tilde{b}$, and form $\Lambda$-levels~\citep{jain2007composite}.
In our case, the gap between neighboring $\Lambda$-levels is~\citep{horvathy_non-commutative_2002,xiao_berry_2005}
\begin{equation}
\Delta=\frac{\hbar e\vert\tilde{b}\vert}{Dm^{\prime}}.\label{eq:gap}
\end{equation}
The presence of $D$ in the denominator is notable. The particle-hole
symmetry requires that the gap at fillings $\nu$ and $1-\nu$ should
be equal when one fixes the external magnetic field and changes the
density~\citep{jain2007composite}. However, if $m^{\prime}(1-\nu)=m^{\prime}(\nu)$
is assumed as in the HLR theory, the gap predicted by Eq.~(\ref{eq:gap})
will be asymmetric for particles and holes since $D=b/B=2\nu$ is
proportional to the filling factor.

To compensate, we explicitly include a factor $D$ to the CF dispersion
in Eq.~(\ref{eq:dispersion}), and interpret $m^{\ast}$ as the effective
mass of CFs. With the factor, the particle-hole symmetry requires
\begin{equation}
m^{*}(\nu)=m^{*}(1-\nu).\label{eq:msymmetry}
\end{equation}
We note that $m^{\ast}$ also depends on $B$, as shown in Ref.~\citep{halperin_theory_1993}.

\section{Intrinsic response\label{sec:Intrinsic-responses}}

\subsection{Intrinsic Hall conductance\label{subsec:Intrinsic-Hall-conductance}}

We first determine the intrinsic response, which is the EM response
when the system is in a local equilibrium state. In our case, the
distribution function of the local equilibrium state is
\begin{equation}
f_{0}\left(\bm{x},\bm{p},t\right)=n_{\mathrm{F}}\left(\mathcal{E}(\bm{p})-\mu\right),\label{eq:f0}
\end{equation}
where $n_{\mathrm{F}}$ is the Fermi-Dirac distribution function,
$\mu$ is the chemical potential, $\mathcal{E}(\bm{p})$ is the energy
dispersion shown in Eq.~(\ref{eq:Epsilon}). The intrinsic contribution
to the current is
\begin{equation}
\bm{j}_{0}=-e\int\left[\mathrm{d}\bm{p}\right]D\dot{\bm{x}}f_{0}\left(\bm{p}\right)\label{eq:j0}
\end{equation}
with $[\mathrm{d}\bm{p}]\equiv\mathrm{d}^{2}p/h^{2}$.

To evaluate the intrinsic current, we note that the electric field
is to shift the center of the CF dispersion, i.e., $\mathcal{E}(\bm{p})\approx\varepsilon\left(\bm{p}-m^{\ast}\bm{E}\times\hat{z}/b\right)$
to the linear order of $\bm{E}$. Substituting Eq.~(\ref{eq:EOM3_1})
and the dispersion into Eq.~(\ref{eq:j0}), and making the substitution
$\bm{p}\rightarrow\bm{p}+m^{\ast}\bm{E}\times\hat{z}/b$, we have
\begin{equation}
\bm{j}_{0}=-e\int[d\bm{p}]Dn_{\mathrm{F}}\left(\varepsilon(\bm{p})-\mu\right)\left(\frac{\bm{p}}{m^{*}}-\frac{\hat{z}\times\tilde{\bm{e}}}{b}\right).
\end{equation}
The first term of the integral vanishes, and the second term can be
determined by using the identity $-e\int[d\bm{p}]Dn_{\mathrm{F}}\left(\varepsilon(\bm{p})-\mu\right)=\rho$
and the self-consistent condition Eq.~(\ref{eq:btorho}). We obtain:
\begin{align}
\bm{j}_{0}= & -\sigma_{\mathrm{CS}}\tilde{\bm{e}}.\label{eq:j02-1}
\end{align}

The intrinsic response contributes a Hall conductance $\tilde{\sigma}_{xy}^{0}=-e^{2}/2h$
for the effective electric field. It is notable that the Hall conductance
is independent of the density or the filling factor.

\subsection{Definition of the CF coordinate}

In the last subsection, we obtain a half-quantized intrinsic Hall
conductance for CFs. It seems to provide a solution to the issue raised
by Kivelson et~al., i.e., CFs must have a half-quantized Hall conductance
in a particle-hole symmetric LL~\citep{kivelson_composite-fermion_1997}.
However, the issue is not fully resolved due to an ambiguity in defining
the coordinate of a CF. The interpretation of the dynamics, in particular
the presence of a Berry curvature, depends on the definition we adopt
for the CF coordinate, while the intrinsic Hall conductance is directly
related to the Berry curvature.

To see that, we examine the Lagrangian Eq.~(\ref{eq:lagrangian})
to see how the definition of the CF coordinate affects its interpretation.
At half-filling, the Lagrangian for a CF has the form $L=\bm{p}\cdot\dot{\bm{x}}+\left(\bm{p}\times\dot{\bm{p}}\right)\cdot\hat{z}/2eB+e\tilde{\phi}(\bm{x})-\mathcal{E}$,
where $\bm{x}$ is the coordinate of the vortex in the CF. We implicitly
define the coordinate of the CF as the coordinate of its vortex and
interpret the second term as a Berry curvature $\Omega_{z}=1/eB$.
However, the interpretation changes with an alternative definition
of the CF coordinate. For instance, if we define the CF coordinate
as the center coordinate $\bm{x}^{\mathrm{c}}=\bm{x}+\bm{d}/2=\bm{x}+\bm{p}\times\hat{z}/2eB$,
the Lagrangian becomes $L=\bm{p}\cdot\dot{\bm{x}}^{\mathrm{c}}+e\tilde{\phi}(\bm{x})-\mathcal{E}$,
and the Berry curvature vanishes. Similarly, if we define the CF coordinate
as the coordinate of its electron, the Berry curvature will reverse
its sign. Then, what would be the physical interpretation?

The answer lies in the local equilibrium state that the system adapts
to. When external fields are applied, the system rapidly relaxes to
a local equilibrium state with its momentum distribution determined
by momentum relaxation processes. By assuming the local equilibrium
distribution Eq.~(\ref{eq:f0}), we implicitly assume that the momentum
relaxation processes do not change $\bm{x}$, i.e., the position of
the vortex, otherwise the effective potential $\tilde{\phi}(\bm{x})$
would affect the energy conservation of relaxation processes and therefore
the detailed balance condition for determining the local equilibrium
distribution. It is the momentum relaxation process instead of an
arbitrary choice of the definition of the CF coordinate that determines
the intrinsic response.

To elucidate the point, we try an alternative definition of the CF
coordinate, say, the center of the dipole $\bm{x}^{\mathrm{c}}$.
In this case, multipole expansions lead to a dispersion $\mathcal{E}^{\prime}(\bm{p})=\varepsilon(\bm{p})-[(\bm{E}-\bm{e})\times\hat{z}]\cdot\bm{p}/2B$.
We still assume that a momentum relaxation process does not change
the position of the vortex. It implies that a process inducing a momentum
change $\Delta\bm{p}$ also induces a change of the CF coordinate
(side jump) $\Delta\bm{x}^{\mathrm{c}}=\Delta\bm{p}\times\hat{z}/2eB$,
and a change of energy $\Delta\mathcal{E}^{\prime}(\bm{p})+e(\bm{e}+\bm{E})\cdot\Delta\bm{x}^{\mathrm{c}}=\Delta\varepsilon(\bm{p})-(\bm{E}\times\hat{z})\cdot\Delta\bm{p}/B\equiv\Delta\mathcal{E}(\bm{p})$.
The principle of detailed balance will lead to the same local equilibrium
distribution Eq.~(\ref{eq:f0}).

In the next section, we will establish that the momentum relaxation
processes in a LL with the particle-hole symmetry do keep $\bm{x}$
unchanged. It makes $\bm{x}$ a convenient choice as the CF coordinate
since one needs not to deal with the complexities associated with
side jumps~\citep{wang_particle-hole_2017}. When the filling factor
deviates from $1/2$, there is another reason for choosing $\bm{x}$
as the CF coordinate. In this case, the particular choice provides
the simplest form of the equations of motions. With other choices,
non-diagonal Berry curvature components will appear, making the equations
more complicated and their interpretation more difficult. The latter
point is discussed in Ref.~\citep{shi_dynamics_2018}.

\section{Scattering of CFs\label{sec:Scattering of CF}}

In this section, we investigate the momentum relaxation processes
of CFs. At low temperatures, the processes are dominated by elastic
scattering induced by quenched impurities. For CF systems, there are
two possible ways that impurities can affect CFs: (a) impurities induce
a random potential coupling to the electrons in dipoles; (b) the random
impurity potential induces a spatial modulation of the CF density,
and in turn induces a spatially fluctuating CS magnetic field. We
will show that CFs undergo no side-jumps when scattered by both the
fields as long as CFs adopt a dispersion consistent to the particle-hole
symmetry requirement Eq~(\ref{eq:msymmetry}). It justifies our choice
of defining the coordinate of a CF as its vortex position.

We will first consider scattering by the impurity potential only.
The more complete consideration including the fluctuating CS field
will be presented in Sec.~\ref{subsec:CS magnetic fluctuations}.
Our consideration is similar to Wang et~al.'s consideration for the
HLR theory~\citep{wang_particle-hole_2017}. The correspondence between
the two considerations will be discussed in Sec.~\ref{subsec:Compared-to-Wang}.
Our result also coincides with numerical observations by Geraedts
et~al.~\citep{geraedts_berry_2018}. It will be discussed in Sec.~\ref{subsec:Compared-to-Geraedts}.

\subsection{Scattering by the impurity potential}

At low temperatures, the momentum relaxation is dominated by elastic
scattering induced by impurities ubiquitously presented in real materials.
Impurities introduce a random potential into the system. As an electrostatic
potential, it only couples to the electrons in dipoles. Electrons
feel an effective random potential $V_{\textrm{eff}}(\bm{x}^{\mathrm{e}})$
which consists of the random potential imposed directly by impurities
and a screening potential from other electrons. It is important to
note that the potential depends on $\bm{x}^{\mathrm{e}}$ instead
of $\bm{x}$.

\begin{figure}[t]
\begin{centering}
\includegraphics{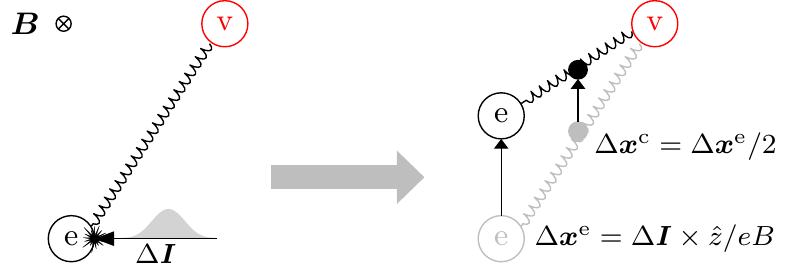}
\par\end{centering}
\caption{\label{figure1} Collision between a CF and an impurity. Left: the
impurity exerts an impulse $\Delta\bm{I}$ to the electron in the
CF. Right: after the collision, the electron coordinate as well as
the center coordinate $\bm{x}^{\mathrm{c}}$ are shifted. Depending
on the definition of the CF coordinate as the center coordinate $\bm{x}^{\mathrm{c}}$
or the vortex coordinate $\bm{x}$, the scattering can be interpreted
as with or without a side jump, respectively.}
\end{figure}

We first present a heuristic picture for a CF colliding with an impurity.
We assume that the scattering is instantaneous, i.e., the impurity
exerts an impulsive force to the electron with a total impulse $\Delta\bm{I}$.
The motion of an electron confined in a LL is governed by the equation
$eB\hat{z}\times\dot{\bm{x}}^{\textrm{e}}=\bm{F}$, where $\bm{F}$
is the total force acting on the electron, including the binding force
from the vortex and the impulsive force exerted by the impurity. Since
the binding force is negligible in the infinitesimal time period of
the collision, the equation predicts that the electron will have a
coordinate shift $\Delta\bm{x}^{\textrm{e}}=\Delta\bm{I}\times\hat{z}/eB$
after the collision. Meanwhile, the position of the vortex does not
change. The process of the collision is illustrated in Fig.~\ref{figure1}.
According to the dipole picture and Eq.~(\ref{eq:momentum}), it
corresponds to a CF momentum change $\Delta\bm{p}=eB\hat{z}\times\Delta\bm{x}^{\mathrm{e}}=\Delta\bm{I}$
and no change to the CF coordinate. We therefore conclude that the
collision does not induce a side jump. Note that the conclusion is
valid only when we define the coordinate of the CF as its vortex position.
Had we chosen $\bm{x}^{\mathrm{c}}$ as the CF coordinate, for instance,
we would conclude that the collision induces a side jump $\Delta\bm{x}^{\mathrm{c}}=\Delta\bm{I}\times\hat{z}/2eB=\Delta\bm{p}\times\hat{z}/2eB$.

To treat the problem more formally, we follow the approach developed
by Sinitsyn et~al.~\citep{sinitsyn_coordinate_2006}. For a canonical
quantum system, it is shown that the side jump induced by elastic
scattering is
\begin{equation}
\Delta\tilde{\bm{x}}_{\tilde{\bm{p}}^{\prime},\tilde{\bm{p}}}=-\hbar(\partial_{\tilde{\bm{p}}}+\partial_{\tilde{\bm{p}}^{\prime}})\textrm{arg}[V_{\tilde{\bm{p}}^{\prime},\tilde{\bm{p}}}],\label{eq:sj1}
\end{equation}
where $V_{\tilde{\bm{p}}^{\prime},\tilde{\bm{p}}}=\braket{\tilde{\bm{p}}^{\prime}|\hat{V}|\tilde{\bm{p}}}$
is the matrix element of a scattering potential $\hat{V}$ in the
momentum eigenstates $\ket{\tilde{\bm{p}}}$ and $\bra{\tilde{\bm{p}}^{\prime}}$,
and $\textrm{arg}[V_{\tilde{\bm{p}}^{\prime},\tilde{\bm{p}}}]$ denotes
its phase angle. Note that the formula is valid for a canonical system,
i.e., its classical Lagrangian should have the form of $L=\tilde{\bm{p}}\cdot\dot{\tilde{\bm{x}}}-H(\tilde{\bm{x}},\tilde{\bm{p}})$,
where $\tilde{\bm{x}}$ and $\tilde{\bm{p}}$ are the canonical coordinate
and momentum, respectively.

For more general Lagrangians in which dynamic variables are not necessarily
canonical, the side jump can be determined by the following procedure:
(a) find a transformation from original dynamical variables $(\bm{x},\bm{p})$
to canonical variables $(\tilde{\bm{x}},\tilde{\bm{p}})$; (b) determine
the scattering matrix element and the side jump $\Delta\tilde{\bm{x}}_{\tilde{\bm{p}}^{\prime},\tilde{\bm{p}}}$
by using Eq.~(\ref{eq:sj1}); (c) determine the side jump in terms
of $(\bm{x},\bm{p})$ by transforming the canonical variables back
to the original dynamic variables.

We can determine the side jump for the current case by applying the
procedure. At half filling, the Lagrangian Eq.~(\ref{eq:lagrangian})
is reduced to $L=\bm{p}\cdot\dot{\bm{x}}+\left(\bm{p}\times\dot{\bm{p}}\right)\cdot\hat{z}/2eB-\varepsilon-V_{\mathrm{eff}}$.
It is easy to find the canonical variables $\tilde{\bm{p}}=\bm{p}$
and $\tilde{\bm{x}}=\bm{x}^{\mathrm{c}}=\bm{x}+\bm{p}\times\hat{z}/2eB$.
Using the canonical variables, the scattering potential can be written
as $V_{\textrm{eff}}(\bm{x}^{\mathrm{e}})=V_{\textrm{eff}}(\tilde{\bm{x}}+\tilde{\bm{p}}\times\hat{z}/2eB)$.
The scattering matrix element can be determined straightforwardly:
\begin{align}
V_{\tilde{\bm{p}}^{\prime},\tilde{\bm{p}}} & =\tilde{V}_{\textrm{eff}}(\bm{q})\exp\left[\mathrm{i}\frac{(\tilde{\bm{p}}^{\prime}\times\tilde{\bm{p}})\cdot\hat{z}}{2eB\hbar}\right],
\end{align}
where $\tilde{V}_{\textrm{eff}}(\bm{q})$ is the Fourier transform
of the effective impurity potential, and $\hbar\bm{q}=\tilde{\bm{p}}^{\prime}-\tilde{\bm{p}}\equiv\Delta\bm{p}$
is the change of the momentum induced by the scattering. By applying
Eq.~(\ref{eq:sj1}), we obtain
\begin{equation}
\Delta\tilde{\bm{x}}_{\tilde{\bm{p}}^{\prime},\tilde{\bm{p}}}=\frac{\hbar\bm{q}\times\hat{z}}{2eB}.\label{eq:sjscalar}
\end{equation}
The result is identical to $\Delta\bm{x}^{\mathrm{c}}$ determined
in the heuristic argument. Transforming back to the original variables,
we have 
\begin{equation}
\Delta\bm{x}_{\bm{p}^{\prime},\bm{p}}=\Delta\tilde{\bm{x}}_{\bm{p}^{\prime},\bm{p}}-\Delta\left[\frac{\bm{p}\times\hat{z}}{2eB}\right]=0.\label{eq:sj2}
\end{equation}
It confirms our conclusion from the heuristic argument that scattering
of CFs by impurities does not induce side jumps.

\subsection{Effect of the fluctuating CS magnetic field\label{subsec:CS magnetic fluctuations}}

In the last subsection, we analyze the scattering of CFs by considering
only the effective impurity potential. However, the potential can
also induce a density modulation of CFs. According to Eq.~(\ref{eq:btorho}),
the density modulation will induce a spatially fluctuating CS magnetic
field. Wang et~al. analyze the effect of the fluctuating CS magnetic
field in the HLR theory, and find that it induces side jumps and gives
rise to a half-quantized CF Hall conductance~\citep{wang_particle-hole_2017}.
Here, we consider the same effect in the dipole picture.

We first determine the modification to the Lagrangian by the fluctuating
CS magnetic field. Following Wang et~al.'s consideration, we can
determine the fluctuating CS magnetic field $\delta b(\bm{x})=-(2m^{*}/e\hbar)V_{\textrm{eff}}(\bm{x})$~\citep{wang_particle-hole_2017}.
It gives rise to a CS vector potential $\delta\bm{a}(\bm{x})$. The
Lagrangian is modified to
\begin{equation}
L=\bm{p}\cdot\dot{\bm{x}}+\frac{1}{2eB}\left(\bm{p}\times\dot{\bm{p}}\right)\cdot\hat{z}-e\delta\bm{a}(\bm{x})\cdot\dot{\bm{x}}-\varepsilon-V_{\mathrm{eff}}\left(\bm{x}^{\mathrm{e}}\right).
\end{equation}

Following the procedure outlined in the last subsection, we proceed
to find a set of canonical variables $(\tilde{\bm{x}},\tilde{\bm{p}})$
for the modified Lagrangian. Unlike the previous case, the presence
of the CS vector potential makes both $\bm{x}$ and $\bm{p}$ non-canonical.
As in Wang et~al.'s consideration, we assume that the impurity potential
is weak and slowly varying in space. It thus suffices to find a transformation
accurate to the linear order of $\delta\bm{a}$ and its first spatial
gradient. We have~\citep{chang_berry_2008}:%
\begin{align}
\bm{x} & =\tilde{\bm{x}}+\frac{1}{2eB}\hat{z}\times\left[\bm{p}+\frac{1}{2B}(\hat{z}\times\bm{p})\cdot\bm{\nabla}\delta\bm{a}(\tilde{\bm{x}})\right],\label{eq:x2}\\
\bm{p} & =\left[1-\frac{\delta b(\tilde{\bm{x}})}{2B}\right]\tilde{\bm{p}}+e\delta\bm{a}(\tilde{\bm{x}}).\label{eq:p2}
\end{align}
It is straightforward to verify that $L=\tilde{\bm{p}}\cdot\dot{\tilde{\bm{x}}}-\varepsilon-V_{\mathrm{eff}}-\mathrm{d}\left[\delta\bm{a}(\bm{x})\cdot\left(\hat{z}\times\bm{p}\right)/2B\right]/\mathrm{d}t+O[(\delta a)^{2},\nabla^{2}a]$,
i.e., $\tilde{\bm{x}}$ and $\tilde{\bm{p}}$ are indeed canonical.

We can then determine the perturbation to the hamiltonian of the canonical
system. Besides the direct perturbation from $\hat{V}_{\mathrm{eff}}\left(\hat{\bm{x}}_{\mathrm{e}}\right)$,
there is a perturbation to the kinetic energy $\varepsilon\left(\hat{p}\right)=D\hat{p}^{2}/2m^{\ast}$
induced by the CS magnetic field. The perturbation can arise from:
(a) the substitution of the momentum operator Eq.~(\ref{eq:p2});
(b) the modulation of $D(\bm{x})\approx1+\delta b(\tilde{\bm{x}})/B$
induced by the CS magnetic field modulation; (c) the dependence of
$m^{\ast}$ on the filling factor $\nu$ which is also modulated with
the local density. For a particle-hole symmetric LL, the contribution
from (c) disappears in the linear order because $m^{\ast}$ is symmetric
about $\nu=1/2$. Moreover, the extra factor before $\tilde{\bm{p}}$
in Eq.~(\ref{eq:p2}) just cancels the contribution from (b). Therefore,
the total perturbation is:
\begin{equation}
\hat{V}=\frac{e}{2m^{\ast}}\left[\hat{\tilde{\bm{p}}}\cdot\delta\bm{a}\left(\tilde{\bm{x}}\right)+\delta\bm{a}\left(\tilde{\bm{x}}\right)\cdot\hat{\tilde{\bm{p}}}\right]+\hat{V}_{\mathrm{eff}}\left(\hat{\bm{x}}_{\mathrm{e}}\right).
\end{equation}

We proceed to determine the scattering matrix element. The Fourier
transform of $\delta\bm{a}(\tilde{\bm{x}})$ is $\delta\tilde{\bm{a}}(\bm{q})=-\mathrm{i}\delta\tilde{b}(\bm{q})(\hat{z}\times\bm{q}/q^{2})=\mathrm{i}(2m^{\ast}/e\hbar q^{2})(\hat{z}\times\bm{q})\tilde{V}_{\mathrm{eff}}(\bm{q})$.
Thus, the total scattering matrix element is
\begin{equation}
V_{\tilde{\bm{p}}^{\prime},\tilde{\bm{p}}}=\tilde{V}_{\textrm{eff}}(\bm{q})\left[e^{\mathrm{i}\frac{(\tilde{\bm{p}}^{\prime}\times\tilde{\bm{p}})\cdot\hat{z}}{2eB\hbar}}+\frac{2\mathrm{i}(\tilde{\bm{p}}^{\prime}\times\tilde{\bm{p}})\cdot\hat{z}}{\hbar^{2}q^{2}}\right].\label{eq:U2}
\end{equation}
Following Wang et al.'s assumption that $\tilde{V}_{\textrm{eff}}(\bm{q})/q$
does not diverge at $q\rightarrow0$, we determine that the phase
angle of the scattering matrix element is $\mathrm{arg}(V_{\tilde{\bm{p}}^{\prime},\tilde{\bm{p}}})\approx-\hbar^{2}q^{2}/2[(\tilde{\bm{p}}^{\prime}\times\tilde{\bm{p}})\cdot\hat{z}]$
to the linear order of $ql_{B}$. 

Applying Eq.~(\ref{eq:sj1}), we determine the side jump in the canonical
system: 
\begin{equation}
\Delta\tilde{\bm{x}}_{\tilde{\bm{p}}^{\prime},\tilde{\bm{p}}}=\frac{\bm{q}\times\hat{z}}{2k_{\mathrm{F}}^{2}},\label{eq:sjcs}
\end{equation}
for CFs on the Fermi circle. Substituting $k_{\mathrm{F}}=\sqrt{eB/\hbar}$
into the relation, we find that it is identical to Eq.~(\ref{eq:sjscalar})
determined for a scalar potential.

Finally, we determine the side-jump in terms of $\bm{x}$ and $\bm{p}$.
Note that the side jump is determined by comparing particle trajectories
in asymptotic regions far from an impurity~\citep{sinitsyn_coordinate_2006}.
In these regions, the CS field $\delta\bm{a}(\bm{x})$ induced by
the impurity is negligible. Therefore, the relation between $(\Delta\bm{x},\Delta\bm{p})$
and $(\Delta\tilde{\bm{x}},\Delta\tilde{\bm{p}})$ is the same as
Eq.~(\ref{eq:sj2}). We thus have
\begin{align}
\Delta\bm{x}_{\bm{p}^{\prime},\bm{p}} & =0.\label{eq:Dx2}
\end{align}
It indicates that when both the impurity potential and the fluctuating
CS magnetic field are considered, the scattering of CFs does not induce
side jumps in a LL with the particle-hole symmetry.

\subsection{Compared to Wang et~al.'s result\label{subsec:Compared-to-Wang}}

The discussion in the last subsection closely follows Wang et~al.'s
model on how impurities affect a CF system. An obvious difference
between the two considerations is in the choice of the CF coordinate:
Wang et~al. adopt $\bm{x}^{\mathrm{c}}$ (or $\tilde{\bm{x}}$) as
the coordinate of a CF, while we use $\bm{x}$. The results obtained
in the two considerations are identical: Eq.~(\ref{eq:sjcs}) is
exactly the side jump obtained by Wang et~al.~\citep{wang_particle-hole_2017}.

In the microscopic level, the presence of the Berry curvature and
the internal structure of CFs in the dipole picture do introduce differences.
In our theory, the impurity potential are coupled to electrons. It
leads to a phase factor in the first term of Eq.~(\ref{eq:U2}).
It is inconsequential because its contribution to the phase angle
of the scattering matrix element is of the higher order in $ql_{B}$.
The more fundamental difference is in Eq.~(\ref{eq:p2}), which is
different from the usual Peierls substitution by a factor $\sim1/\sqrt{D}$.
It is also inconsequential because it is cancelled by the $D$ factor
in the CF dispersion Eq.~(\ref{eq:dispersion}). The resulting perturbation
Eq.~(\ref{eq:U2}) is identical to that of Wang's theory except for
the aforementioned inconsequential difference in the effective impurity
potential.

\subsection{Compared to Geraedts et al.'s result\label{subsec:Compared-to-Geraedts}}

As an attempt to determine the Berry phase and Berry curvature of
CFs, Geraedts et~al. numerically calculate the matrix element of
the density operator $\hat{\rho}_{-\bm{q}}=\sum_{a}\exp(\mathrm{i}\bm{q}\cdot\bm{r}_{a})$
between two Rezayi-Read states with different sets of  wavevectors~\citep{geraedts_berry_2018}.
In Ref~\citep{ji_berry_2020}, we argue that their calculation could
be regarded as a ``first-principles'' determination of the scattering
matrix element from the microscopic wave function. It is then interesting
to see how the scattering matrix element Eq.~(\ref{eq:U2}) determined
from an effective picture is compared to the ``first-principles''
result.

We find that the simple formula Eq.~(\ref{eq:U2}) captures all essential
features observed in Geraedts et~al.'s calculation. To see that,
we  rewrite Eq.~(\ref{eq:U2}) in the limit $\bm{q}\rightarrow0$
as
\begin{equation}
V_{\tilde{\bm{p}}^{\prime},\tilde{\bm{p}}}\approx-\mathrm{i}\frac{2\tilde{V}_{\textrm{eff}}(\bm{q})}{\hbar q}\tilde{p}\sin\theta\exp\left(\mathrm{i}\frac{\hbar q}{2\tilde{p}\sin\theta}\right)
\end{equation}
where $\theta$ denotes the polar angle from $\tilde{\bm{p}}$ to
$\bm{q}$, and $\tilde{V}_{\textrm{eff}}(\bm{q})$ in the current
context is the effective potential induced by a single-mode scalar
potential $V(\bm{r})=\exp\left(\mathrm{i}\bm{q}\cdot\bm{r}\right)$.
The formula predicts that (a) when $\bm{q}$ is parallel to $\tilde{\bm{p}}$,
i.e., $\theta=0$, the matrix element vanishes; (b) when $\bm{q}$
is perpendicular to $\tilde{\bm{p}}$, i.e., $\theta=\pm\pi/2$, the
matrix element has a phase equal to $\pm(\hbar q/2\tilde{p}-\pi/2)$;
(c) traversing anti-clockwise an circular path around the origin of
the momentum space and cumulating the first part of the phase, one
always obtains a total phase $\pi$, independent of the radius of
the path. These are exactly what are observed in Geraedts et~al.'s
work.

The phase obtained from the scattering matrix element is in general
not the Berry phase~\citep{ji_berry_2020}. Geraedts et~al. interprets
the lack of dependence of the cumulated phase on the radius of the
circular path as a manifestation of the singular distribution of the
Berry curvature in a massless Dirac cone. However, we see that it
is also a property of the scattering matrix element Eq.~(\ref{eq:U2})
even though it is for a system with a uniform Berry curvature.

Nonetheless, Geraedts et al.'s result does provide a ``first principles''
support for the model of impurity scattering shown in Sec.~\ref{subsec:CS magnetic fluctuations}.

\section{Extrinsic response \label{sec:extrinsic responses}}

\subsection{Boltzmann equation}

In this section, we determine the extrinsic response of CFs. It is
the contribution from the deviation of the non-equilibrium distribution
of the system from the local equilibrium distribution Eq.~(\ref{eq:f0}).
The non-equilibrium distribution function $f=f(t,\bm{x},\bm{p})$
is determined by the Boltzmann equation:
\begin{gather}
\partial_{t}f+\dot{\bm{x}}\cdot\left(\partial_{\bm{x}}f\right)+\dot{\bm{p}}\cdot\left(\partial_{\bm{p}}f\right)=\left.\frac{\partial f}{\partial t}\right|_{\textrm{coll}},\label{eq:BE}
\end{gather}
where $\left.\partial f/\partial t\right|_{\textrm{coll}}$ denotes
the collision contribution of impurity scattering.

The collision contribution can be explicitly specified. In the last
section, we show that CFs undergo no side jumps when scattered. It
greatly simplifies our consideration because we can model the collision
term as usual for elastic scattering~\citep{callaway1991quantum}:
\begin{equation}
\left.\frac{\partial f}{\partial t}\right|_{\textrm{coll}}=-\int[d\bm{p}^{\prime}]DW_{\bm{p},\bm{p}^{\prime}}[f(t,\bm{x},\bm{p})-f(t,\bm{x},\bm{p}^{\prime})],\label{eq:collision}
\end{equation}
where $W_{\bm{p},\bm{p}^{\prime}}$ denotes the transition probability
from state $\bm{p}^{\prime}$ to $\bm{p}$. For elastic scattering,
the energy conservation requires $W_{\bm{p},\bm{p}^{\prime}}\propto\delta[\mathcal{E}(\bm{p})-\mathcal{E}(\bm{p}^{\prime})]$.

The solution of the Boltzmann equation can be decomposed into
\begin{equation}
f(t,\bm{x},\bm{p})=f_{0}(t,\bm{x},\bm{p})+f_{1}(t,\bm{x},\bm{p}),
\end{equation}
where $f_{0}$ is the local equilibrium distribution specified in
Eq.~(\ref{eq:f0}), and $f_{1}$ is the deviation induced by driving
fields. It is obvious that $f_{0}$ fulfills the condition of detailed
balance, i.e., the collision term vanishes for $f_{0}$.

The collision term can be further simplified. We assume that the scattering
is approximately isotropic, $W_{\bm{p},\bm{p}^{\prime}}=W\delta[\mathcal{E}(\bm{p})-\mathcal{E}(\bm{p}^{\prime})]$.
By substituting it into the collision term, and noting that $\mathcal{E}(\bm{p})$
and $\mathcal{E}(\bm{p}^{\prime})$ can be replaced with $\varepsilon(\bm{p})$
and $\varepsilon(\bm{p}^{\prime})$ since $f_{1}$ has already been
in the linear order of driving fields, we have~\citep{mirlin_composite_1997}
\begin{gather}
\left.\frac{\partial f}{\partial t}\right|_{\textrm{coll}}=-\frac{f_{1}-\bar{f}_{1}}{\tau},\label{eq:relaxation time approximation}
\end{gather}
where $1/\tau=\int[d\bm{p}^{\prime}]DW\delta(\varepsilon_{\bm{p}}-\varepsilon_{\bm{p}^{\prime}})$
is the reciprocal of the relaxation time, and 
\begin{equation}
\bar{f}_{1}=\frac{1}{2\pi}\int_{0}^{2\pi}f_{1}(\phi)d\phi
\end{equation}
is the average of $f_{1}$ over the polar angle $\phi$ of the momentum
$\bm{p}$. Strictly speaking, the transition probability for our case
must be anisotropic, as evident in Eq.~(\ref{eq:U2}). However, in
the regime $ql_{B}\ll1$ we are interested in, the anisotropy will
not introduce qualitative differences. Therefore, we ignore the anisotropy
for simplicity. The generalization for anisotropic scattering is straightforward
and can be found in Ref.~\citep{mirlin_composite_1997}.

To solve the Boltzmann equation to the linear order of driving fields,
it suffices to solve for single-mode fields $[\bm{E}(t,\bm{x}),\,\bm{e}(t,\bm{x})]=(\bm{E},\,\bm{e})\exp(-\mathrm{i}\omega t+\mathrm{i}\bm{q}\cdot\bm{x})$.
The solution has the form $f_{1}(t,\bm{x},\bm{p})=f_{1}\exp(-\mathrm{i}\omega t+\mathrm{i}\bm{q}\cdot\bm{x})$.
By substituting Eqs.~(\ref{eq:EOM3_1},~\ref{eq:EOM3_2}) into Eq.~(\ref{eq:BE}),
we have:
\begin{multline}
\left[-1/\tau+\mathrm{i}\omega-\mathrm{i}\bm{v}\cdot\bm{q}-\omega_{\textrm{c}}^{*}(\hat{z}\times\bm{p})\cdot\partial_{\bm{p}}\right]f_{1}\\
=e\bm{v}\cdot(\bm{e}+\beta\bm{E})\left(-\frac{\partial f_{0}}{\partial\varepsilon}\right)-\frac{\bar{f}_{1}}{\tau},\label{eq:BEsinglemode}
\end{multline}
where $\bm{v}=\bm{p}/m^{\ast}$, $\beta$ is defined in Eq.~(\ref{eq:beta}),
and $\omega_{\textrm{c}}^{*}=e\tilde{b}/m^{*}$ is the effective CF
cyclotron frequency. The solutions of the equation for $\nu=1/2$
and $\nu\ne1/2$ will be discussed in the following subsections.

With the distribution function, the extrinsic contribution to the
CF (vortex) current density can be determined by
\begin{equation}
\bm{j}_{1}=-e\int[d\bm{p}]Df_{1}\dot{\bm{x}}\approx-e\int[d\bm{p}]Df_{1}\frac{\bm{p}}{m}.\label{eq:j1}
\end{equation}

\subsection{$\nu=1/2$\label{sec:CF-current}}

We first consider the case of half-filling. In this case, the effective
magnetic field $\tilde{b}=0$, therefore $\omega_{\mathrm{c}}^{\ast}=0$.
Equation~(\ref{eq:BEsinglemode}) can be readily solved. We have
\begin{align}
f_{1} & =\frac{e\bm{v}\cdot(\bm{e}+\beta\bm{E})\left(-\frac{\partial f_{0}}{\partial\varepsilon}\right)-\bar{f}_{1}/\tau}{-1/\tau+\mathrm{i}\omega-\mathrm{i}\bm{q}\cdot\bm{v}}.\label{eq:f1}
\end{align}
By averaging both sides of Eq.~(\ref{eq:f1}) over the polar angle
of the momentum, we obtain a self-consistent equation for determining
$\bar{f}_{1}$. After solving the equation, we have
\begin{gather}
\bar{f}_{1}=\tau e\bar{\bm{v}}\cdot(\bm{e}+\beta\bm{E})\left(-\frac{\partial f_{0}}{\partial\varepsilon}\right),
\end{gather}
where $\bar{\bm{v}}=-iv_{\textrm{F}}\bm{q}l/\{(1-i\omega\tau)(1+\gamma)[1-\gamma(1-i\omega\tau)]\}$,
$v_{\textrm{F}}=\hbar k_{\textrm{F}}/m^{*}$ is the CF Fermi velocity,
$l=v_{\textrm{F}}\tau$ is the mean free path of CFs, and 
\begin{equation}
\gamma=\sqrt{1+\frac{(ql)^{2}}{(1-i\omega\tau)^{2}}}.\label{eq:gamma}
\end{equation}
The solution of the Boltzmann equation can be written as
\begin{align}
f_{1} & =\frac{e(\bm{v}-\bar{\bm{v}})\cdot(\bm{e}+\beta\bm{E})}{-1/\tau+\mathrm{i}\omega-\mathrm{i}\bm{q}\cdot\bm{v}}\left(-\frac{\partial f_{0}}{\partial\varepsilon}\right).\label{eq:f1-1}
\end{align}

Substituting Eq.~(\ref{eq:f1-1}) into Eq.~(\ref{eq:j1}), we obtain
\begin{align}
\bm{j}_{1} & =\tilde{\sigma}(\bm{e}+\beta\bm{E}),\label{eq:j1response}\\
\tilde{\sigma}_{11} & =\frac{2\sigma_{0}}{(1-\mathrm{i}\omega\tau)\left(1+\gamma\right)}\frac{\mathrm{i}\omega\tau}{1-\gamma(1-\mathrm{i}\omega\tau)},\\
\tilde{\sigma}_{22} & =\frac{2\sigma_{0}}{(1-\mathrm{i}\omega\tau)(1+\gamma)},\label{eq:sigma22}
\end{align}
where $\sigma_{0}=e^{2}n\tau/m^{*}$ is the Drude conductivity, $n$
is the density of CFs, and the subscripts $1$ and $2$ denote the
longitudinal ($\parallel\bm{q}$) and transverse ($\parallel\hat{z}\times\bm{q}$)
directions, respectively. The non-diagonal elements of $\tilde{\sigma}$
vanish at half-filling.

It is important to note that the extrinsic response is not driven
by the effective electric field $\tilde{\bm{e}}$ but by $\bm{e}+\beta\bm{E}$.
As a result, $\bm{E}$ is not equivalent to the CS electric field
$\bm{e}$ in driving the response. It manifests the fact that CFs
in the dipole picture are not point particles and subject to the dipole
correction from which the $\beta$ matrix arises. The dipole correction
is derived from the internal structure of CFs. This is in contrast
with the Dirac CF theory, in which a CF is regarded as a point particle
and the dipole moment is introduced only as an after-thought for fulfilling
the requirement of the Galilean invariance~\citep{son_is_2015,nguyen2018}.

\subsection{$\nu\protect\neq1/2$\label{sec:Conductivity-relation-away}}

Away from half filling, the effective magnetic field $\tilde{b}=b-B$
is nonzero. Solving Eq.~(\ref{eq:BEsinglemode}) becomes more complicated.
To this end, we note that Eq.~(\ref{eq:BEsinglemode}) is identical
to the Boltzmann equation for ordinary particles albeit with a driving
field $\bm{e}+\beta\bm{E}$. For ordinary systems, Mirlin and W$\ddot{\textrm{o}}$lfle
obtain a solution in Ref.~\citep{mirlin_composite_1997}. Their solution
can be adapted to our case by noting the relation $-f_{0}^{\prime}\left(\varepsilon(\bm{p})\right)=D^{-1}\delta\left((\bm{p}^{2}-p_{\mathrm{F}}^{2})/2m^{\ast}\right)$.
We obtain
\begin{align}
\tilde{\sigma}_{11} & =-2\tilde{\sigma}_{0}\frac{\mathrm{i}\omega\tau}{(ql)^{2}}\left(1-\mathrm{i}\omega\tau\frac{A_{00}}{A_{00}-\omega_{\textrm{c}}^{*}\tau}\right),\\
\tilde{\sigma}_{22} & =\frac{2\tilde{\sigma}_{0}}{\omega_{\textrm{c}}^{*}\tau}\left(\frac{A_{0s}^{2}}{A_{00}-\omega_{\textrm{c}}^{*}\tau}+A_{ss}\right),\\
\tilde{\sigma}_{12} & =-\tilde{\sigma}_{21}=-2\tilde{\sigma}_{0}\frac{\omega}{v_{\textrm{F}}q}\frac{A_{0s}}{A_{00}-\omega_{\textrm{c}}^{*}\tau},
\end{align}
where $\tilde{\sigma}_{0}=\sigma_{0}/D\equiv\tilde{n}e^{2}\tau/m^{\ast}$is
the Drude conductivity for a system with the effective density $\tilde{n}=n/D=eB/2h$,
and the coefficients $A_{ij}$ are defined by:
\begin{gather}
A_{ij}=\frac{1}{2\pi}\int_{0}^{2\pi}d\phi\psi_{i}(\phi)\int_{\mp\infty}^{\phi}d\phi^{\prime}\psi_{j}(\phi^{\prime})\exp\{K(\phi,\phi^{\prime})\},\nonumber \\
K(\phi,\phi^{\prime})=-\frac{\phi-\phi^{\prime}}{\omega_{\textrm{c}}^{*}\tau}(1-\mathrm{i}\omega\tau)-\frac{\mathrm{i}qv_{\textrm{F}}}{\omega_{\textrm{c}}^{*}}(\sin\phi-\sin\phi^{\prime}),\nonumber \\
\psi_{0}(\phi)=1,\quad\psi_{s}(\phi)=\sin\phi,
\end{gather}
where the lower bound of the integral over $\phi^{\prime}$ is $-\infty$
($\infty$) for $\tilde{b}>0$ ($\tilde{b}<0$).

The result has an important feature: it is identical to the response
of a system with an effective density $\tilde{n}=eB/2h$ and a Fermi
wavevector $k_{\mathrm{F}}=\sqrt{4\pi\tilde{n}}=\sqrt{eB/\hbar}$,
which are exactly the parameters of a Dirac CF system~\citep{son_is_2015}.
The effective density is set by the magnetic field instead of the
electron density, exactly the swap of notions observed in the Dirac
CF theory.

The formula hides a subtle difference from the HLR theory which could
have an experimental effect. At finite $q$, the formula predicts
a magneto-conductivity oscillation with respect to $qv_{\mathrm{F}}/|\omega_{\mathrm{c}}^{\ast}|=qR_{\mathrm{c}}^{\ast}$,
where $R_{\mathrm{c}}^{\ast}=\hbar k_{\mathrm{F}}/e|\tilde{b}|$ is
the effective cyclotron radius of CFs~\citep{halperin_theory_1993,mirlin_composite_1997,willet1998}
(see Sec.~\ref{sec:An-experimental-test}). Because the Fermi wavevector
is now determined by the magnetic field, the magneto-conductivity
oscillation will show asymmetry about $\nu=1/2$ in the positions
of conductivity extrema~\citep{ji_asymmetry_2020}, similar to that
observed in recent geometric resonance experiments~\citep{kamburov_composite_2013,kamburov_what_2014}.

\section{Electromagnetic response\label{sec:Electromagnetic-responses}}

\subsection{CF current\label{subsec:CF-current}}

Combining the intrinsic response Eq.~(\ref{eq:j02-1}) and the extrinsic
response Eq.~(\ref{eq:j1response}) of the CF current, we have
\begin{equation}
\bm{j}=-\sigma_{\mathrm{CS}}(\bm{e}+\bm{E})+\tilde{\sigma}\left(\bm{e}+\beta\bm{E}\right).\label{eq:jcf}
\end{equation}
It is important to note that $\tilde{\sigma}$ is a Fermi surface
property. As we have shown in the last section, it is identical to
that of an effective Dirac CF system.

To determine the actual response to external EM fields, we apply the
self-consistent condition Eq.~(\ref{eq:jtoe}), and find
\begin{equation}
\bm{e}=\left(\tilde{\sigma}^{-1}\sigma_{\mathrm{CS}}-\beta\right)\bm{E}.\label{eq:etoE}
\end{equation}
Substituting the relation into Eq.~(\ref{eq:jtoe}), we obtain
\begin{equation}
\bm{j}=\left(-\sigma_{\mathrm{CS}}\tilde{\sigma}^{-1}\sigma_{\mathrm{CS}}+\sigma_{\mathrm{CS}}\beta\right)\bm{E}.\label{eq:jtoE}
\end{equation}
It yields a conductivity tensor exactly the form of Eq.~(\ref{eq:DiracSigmaTrans}).

\subsection{Emergence of the Dirac CF theory\label{subsec:Emergence-of-the}}

To elucidate how a Dirac CF-like EM response emerges in the dipole
picture, we identify the extrinsic contribution $\bm{j}_{1}=\bm{j}-\bm{j}_{0}=\bm{j}+\sigma_{\mathrm{CS}}\tilde{\bm{e}}$
as the ``Dirac CF'' current $\tilde{\bm{j}}$, and $\bm{j}_{\mathrm{D}}=\bm{j}-\sigma_{\mathrm{CS}}\bm{E}$
as the ``Dirac electron'' current~\citep{son_is_2015}. By applying
the equations of continuity for CF and electron densities, we can
generalize them to the densities $\tilde{\rho}=\rho+(e^{2}/2h)\tilde{b}$
and $\rho_{\mathrm{D}}=\rho+e^{2}B/2h$, respectively. We thus have
$j_{\mathrm{D}}^{\mu}=j^{\mu}+(e^{2}/2h)\epsilon^{\mu\nu\gamma}\partial_{\nu}A_{\gamma}$.
Substituting the CS self-consistent conditions Eqs.~(\ref{eq:btorho},~\ref{eq:jtoe}),
or $j^{\mu}=(e^{2}/2h)\epsilon^{\mu\nu\gamma}\partial_{\nu}a_{\gamma}$
into the relation, we obtain
\begin{align}
j_{\mathrm{D}}^{\mu} & =\frac{e^{2}}{2h}\epsilon^{\mu\nu\gamma}\partial_{\nu}\tilde{a}_{\nu}=j_{0}^{\mu}.\label{eq:jD}
\end{align}
We find that the ``Dirac electron'' current is nothing but the intrinsic
current determined in Eq.~(\ref{eq:j02-1}). Substituting the relation
and the CS self-consistent conditions into the identity $\tilde{j}=j-j_{0}$,
We obtain
\begin{equation}
\tilde{j}^{\mu}=-\frac{e^{2}}{2h}\epsilon^{\mu\nu\gamma}\partial_{\nu}A_{\gamma}.\label{eq:jtilde}
\end{equation}
The two relations Eqs.~(\ref{eq:jD},~\ref{eq:jtilde}) are exactly
the self-consistent conditions yielded by the action of the Dirac
CF theory~\citep{son_is_2015}. 

We can rewrite the action Eq.~(\ref{eq:Sfirstorder}) in terms of
$\tilde{j}$ and the effective fields $\tilde{a}$, we have
\begin{multline}
S=\int\mathrm{d}^{2}\bm{x}\mathrm{d}t\left[\frac{e^{2}}{4h}\epsilon^{\mu\nu\gamma}A_{\mu}\partial_{\nu}A_{\gamma}-\frac{e^{2}}{2h}\epsilon^{\mu\nu\gamma}\tilde{a}_{\mu}\partial_{\nu}A_{\gamma}\right.\\
\left.-\tilde{a}\cdot\tilde{j}+L_{\mathrm{Dipoles}}\right].\label{eq:SDirac}
\end{multline}
It has the same gauge structure as the Dirac CF action, although its
particles are governed by a different Lagrangian $L_{\mathrm{Dipoles}}=\bm{M}\cdot\bm{B}-\mathcal{E}-\mathcal{E}_{\mathrm{0}}$
with $\mathcal{E}_{\mathrm{0}}=\tilde{a}[j_{\mathrm{D}}]\cdot j_{\mathrm{D}}/2$
and $\tilde{a}[j_{\mathrm{D}}]$ determined by solving Eq.~(\ref{eq:jD}).
Nonetheless, as far as Fermi-surface properties, e.g., the response
of $\tilde{j}$ to external fields, are concerned, the particles are
indistinguishable from Dirac CFs because they have the same effective
density, Fermi wavevector and Berry phase (see Sec.~\ref{sec:Conductivity-relation-away}).
We thus find that a Dirac CF effective theory emerges from the dipole
picture.

Transforming the action Eq.~(\ref{eq:Sfirstorder}) to Eq.~(\ref{eq:SDirac})
redefines the vacuum of the effective theory from an empty (or filled)
LL of electrons to a half-filled LL of vortices. It is feasible because
the intrinsic response Eq.~(\ref{eq:j02-1}) contributes a half-quantized
Hall conductance which is independent of the filling factor and not
altered by impurity scattering. We can interpret the intrinsic response
as the contribution from the redefined vacuum, $\mathcal{E}_{0}$
as the polarization energy of the vacuum, and the remainder of the
response as from an effective Dirac CF system.

\subsection{Electron current}

The electron current $\bm{j}^{\mathrm{e}}$ is the current actually
measured in experiments. It is related to the CF current $\bm{j}$
by Eq.~(\ref{eq:jeexpansion}). To determine the electron current,
we need to determine $\bm{P}$ and $\bm{M}$.

To determine the dipole density $\bm{P}$, we make use of Eq.~(\ref{eq:EOM3_1}).
Applying the average $-e\int[\mathrm{d}\bm{p}]Df$ to both sides of
the equation, we obtain
\begin{equation}
\bm{j}=\tilde{\omega}_{B}\hat{z}\times\bm{P}-\sigma_{\mathrm{CS}}\bm{e}.
\end{equation}
Applying the self-consistent condition Eq.~(\ref{eq:jtoe}), we find
\begin{equation}
\bm{P}=0.\label{eq:Peq0}
\end{equation}

For the magnetization density $\bm{M}$, it is easy to see that $\bm{M}$
is proportional to $\bm{B}$ and $\bm{b}$. Since we ignore all variations
of $\bm{B}$ and $\bm{b}$ for the moment, we conclude that $\bm{M}$
does not contribute to the response of $\bm{j}^{\mathrm{e}}$.

We thus have
\begin{equation}
\bm{j}^{\mathrm{e}}=\bm{j}
\end{equation}
when only the dipole correction is considered.

\section{Quadrupole and magnetic corrections \label{sec:Quadrupole-and-magnetic}}

In the previous sections, we ignore the quadrupole term in Eq.~(\ref{eq:Epsilon})
as well as the temporal and spatial dependences of the magnetic fields.
In this section, we will consider their effects on the EM response.

The corrections can be classified by orders of $ql_{B}$. The EM response
determined in the last section is accurate only to the zeroth order
of $ql_{B}$. By considering the electric quadrupole correction in
Eq.~(\ref{eq:dispersion}), we can determine the response accurate
to the linear order of $ql_{B}$. Because $\partial_{t}\bm{B}=-\bm{\nabla}\times\bm{E}$
is of the same order of the quadrupole correction, it should be considered
on an equal footing. On the other hand, corrections due to the spatial
gradient of $\bm{B}$ is of the quadratic order in $ql_{B}$, and
considering them requires a multipole expansion of $\bm{j}^{\mathrm{e}}$
to at least the order of the magnetic quadrupole~\citep{jackson1999}.
It is beyond the scope of the current study.

Corrections due to the variation of $\bm{b}$ should be fully accounted
for. Even thought $\bm{b}$ looks like the counterpart of $\bm{B}$,
it is actually equivalent to the density because of the self-consistent
condition Eq.~(\ref{eq:btorho}). Since it is not subject to the
approximation of the multipole expansions, considering its gradients
will not introduce inconsistencies. As we will see, its corrections
are essential for obtaining a density response function of CFs with
a correct static limit.

For simplicity, we consider these corrections only for half filling.

\subsection{Quadrupole correction}

By using Eq.~(\ref{eq:lagrangian}) and the complete form of Eq.~(\ref{eq:Epsilon}),
we determine modifications to the equations of motion Eqs.~(\ref{eq:EOMxdot},
\ref{eq:EOMpdot}) due to the quadrupole correction to the energy
and the time dependence of $B$. We have

\begin{align}
\dot{\bm{x}} & =\frac{b}{B}\frac{\bm{p}}{m^{\ast}}-\frac{1}{eB}\dot{\bm{p}}\times\hat{z}+\frac{1}{eB^{2}}\hat{z}\times\bm{\nabla}\bm{E}\cdot(\bm{p}\times\hat{z}),\label{eq:EOM2_1}\\
\dot{\bm{p}} & =-\frac{1}{B}\bm{\nabla}\bm{E}\cdot(\bm{p}\times\hat{z}),\label{eq:EOM2_2}
\end{align}
where we keep only terms related to $\bm{\nabla}\bm{E}$ since only
they are relevant to the quadrupole correction for the linear response
to EM fields. Substituting Eq.~(\ref{eq:EOM2_2}) into Eq.~(\ref{eq:EOM2_1}),
we obtain $\dot{\bm{x}}=\bm{p}/m^{\ast}$, i.e., the quadrupole correction
does not change Eq.~(\ref{eq:EOM3_1}).

The Boltzmann equation can then be solved. We decompose the distribution
function into $f=f_{0}+f_{1}$, where $f_{0}$ still assumes the same
form as Eq.~(\ref{eq:f0}) but with the quadrupole term in Eq.~(\ref{eq:Epsilon})
included. It is easy to see that it introduces no correction to $\bm{j}_{0}$.
The correction to $f_{1}$ is
\begin{align}
\Delta f_{1} & =\frac{e\left(\bm{v}^{\prime}-\bar{v}^{\prime}\right)\cdot\bm{E}}{-1/\tau+\mathrm{i}\omega-\mathrm{i}\bm{q}\cdot\bm{v}}\left(-\frac{\partial f_{0}}{\partial\varepsilon}\right),\label{eq:f1-2}
\end{align}
where $\bm{v}^{\prime}=-(\omega/2\tilde{\omega}_{B}^{2})\bm{q}\cdot(\hat{z}\times\bm{v})(\hat{z}\times\bm{v})$
and $\bar{\bm{v}}^{\prime}=-v_{\textrm{F}}(\bm{q}l_{B})(\omega/\tilde{\omega}_{B})\gamma/[2(1+\gamma)(1-\gamma(1-i\omega\tau))]$.

The quadrupole correction to $\bm{j}_{1}$ can then be determined
by using Eq.~(\ref{eq:j1}). It is obvious from Eq.~(\ref{eq:Epsilon})
that the quadrupole term has a similar effect as the dipole term in
making $\bm{E}$ nonequivalent to $\bm{e}$. Therefore, the quadrupole
correction can be characterized by a correction to the $\beta$ matrix.
We have

\begin{align}
\Delta\beta_{11} & =-\frac{\left(ql_{B}\right)^{2}}{4}\frac{1+\gamma(1-i\omega\tau)}{(1+\gamma)(1-i\omega\tau)}\approx0,\\
\Delta\beta_{22} & =-\frac{ql_{B}}{4}\frac{\mathrm{i}\omega\left(ql\right)}{\tilde{\omega}_{B}}\frac{1}{(1+\gamma)(1-i\omega\tau)},
\end{align}
and non-diagonal elements vanish. The correction then propagates to
Eq.~(\ref{eq:jtoE}) for the total EM response.

To determine the electron current $\bm{j}^{\mathrm{e}}$, we need
to determine the displacement current $\bm{J}^{\mathrm{Q}}=-\partial\bm{\nabla}\cdot\mathbb{Q}/\partial t$.
Note that $\bm{P}=0$ is still true since Eq.~(\ref{eq:EOM3_1})
is not modified by the quadrupole correction. One may evaluate $\bm{J}^{\mathrm{Q}}$
directly, or apply the Onsager relation which yields in the current
context
\begin{equation}
\bm{J}^{\mathrm{Q}}=\left(\Delta\beta\tilde{\sigma}\right)\left(\bm{e}+\beta\bm{E}\right)=\Delta\beta\sigma_{\mathrm{CS}}\bm{E},
\end{equation}
where we apply Eq.~(\ref{eq:etoE}) to get the last form. The total
correction to the electron conductivity tensor of the system is
\begin{equation}
\Delta\sigma=\sigma_{\mathrm{CS}}\Delta\beta+\Delta\beta\sigma_{\mathrm{CS}},
\end{equation}
where the first term is from the $\Delta\beta$ correction to Eq.~(\ref{eq:jtoE}).
To the linear order of $ql_{B}$, it contributes a correction to the
Hall conductivity
\begin{equation}
\Delta\sigma_{xy}=-\frac{e^{2}}{h}\frac{ql_{B}}{8}\frac{\mathrm{i}\omega}{\tilde{\omega}_{B}}\frac{ql}{(1+\gamma)(1-i\omega\tau)}.
\end{equation}

\subsection{Magnetic correction\label{subsec:Magnetization-current}}

In this subsection, we determine corrections due to the variation
of the CS magnetic field $\delta b$. It has two effects. First, it
modifies the equations of motion, and introduces corrections to the
responses of the CF current. Second, it gives rise to a magnetization
current because the CF dispersion Eq.~(\ref{eq:dispersion}) implies
that each CF carries a magnetic moment $\bm{m}_{b}=-(p^{2}/2m^{\ast}B)\hat{z}$
with respect to the CS magnetic field. Both the effects should be
considered.

First, we determine the corrections to the CF responses. In this case,
the local equilibrium distribution is assumed to be
\begin{equation}
f_{0}\left(\bm{x},\bm{p},t\right)=n_{\mathrm{F}}\left(\varepsilon(\bm{p})-\varepsilon(\bm{p}_{\mathrm{F}})\right),
\end{equation}
where we drop terms in the energy irrelevant to $\delta b$ and assume
a local chemical potential $\mu(\bm{x})=\varepsilon(\bm{p}_{\mathrm{F}})=b(\bm{x})p_{\mathrm{F}}^{2}/2m^{\ast}B$
so that the distribution has a uniform density. Note that the choice
of $\mu(\bm{x})$ does not affect the steady-state solution of the
Boltzmann equation.

We can then determine the correction to the intrinsic current $\bm{j}_{0}$.
$\delta b$ introduces extra forces $-(p^{2}/2m^{\ast}B)\bm{\nabla}\delta b-e\delta b\dot{\bm{x}}\times\hat{z}$
into the system. As a result, Eqs.~(\ref{eq:EOM3_1}, \ref{eq:EOM3_2})
are modified to
\begin{align}
\dot{\bm{x}} & =\frac{\bm{p}}{m^{*}}-\frac{1}{b}\hat{z}\times\bm{e}+\frac{1}{eb}\bm{m}_{b}\times\bm{\nabla}\delta b,\label{eq:xdotdb}\\
\dot{\bm{p}} & =e\delta b\hat{z}\times\frac{\bm{p}}{m^{*}}-e\left(\frac{\bm{e}}{D}+\bm{E}\right)-\frac{\bm{\nabla}\delta b}{b}\frac{p^{2}}{2m^{\ast}},\label{eq:pdotdb}
\end{align}
The correction to $\bm{j}_{0}$ is
\begin{equation}
\Delta\bm{j}_{0}=-\int[\mathrm{d}\bm{p}]Df_{0}\frac{\bm{m}_{b}\times\bm{\nabla}\delta b}{b}=\frac{e^{2}}{16\pi m^{*}}\hat{z}\times\bm{\nabla}\delta b.\label{eq:dj0}
\end{equation}

The correction to the extrinsic current $\bm{j}_{1}$ can also be
determined. It is straightforward to determine that $\delta b$ introduces
a correction to $f_{1}$:
\begin{equation}
\Delta f_{1}=\frac{e(\bm{v}-\bar{\bm{v}})\cdot\left(\frac{\hbar}{2m^{\ast}}\bm{\nabla}\delta b\right)}{-1/\tau+\mathrm{i}\omega-\mathrm{i}\bm{q}\cdot\bm{v}}\left(-\frac{\partial f_{0}}{\partial\varepsilon}\right),
\end{equation}
i.e., the effect of $\delta b$ is equivalent to an effective electric
field $(\hbar/2m)\bm{\nabla}\delta b$. Therefore, the correction
to $\bm{j}_{1}$ is
\begin{equation}
\Delta\bm{j}_{1}=\tilde{\sigma}\frac{\hbar}{2m^{*}}\bm{\nabla}\delta b.\label{eq:dj1}
\end{equation}

Next, we determine the magnetization current. The CS magnetization
is determined by $\bm{M}_{b}=\int[d\bm{p}]Df\bm{m}_{b}.$ We decompose
$\bm{m}_{b}$ as $\bm{m}_{b}(p_{\mathrm{F}})+[\bm{m}_{b}(p)-\bm{m}_{b}(p_{\mathrm{F}})]$,
and note that $f_{1}$ does not contribute to the expectation value
of the second term. We obtain
\begin{align}
\bm{M}_{b} & =\frac{\hbar\rho}{2m^{\ast}}\hat{z}-\int[d\bm{p}]Df_{0}\frac{p^{2}-p_{\mathrm{F}}^{2}}{2m^{\ast}B}\hat{z}=\frac{\hbar\rho}{4m^{\ast}}\hat{z}.
\end{align}
After applying Eq.~(\ref{eq:btorho}), we find that the magnetization
current is
\begin{equation}
\bm{j}^{\mathrm{m}}=\bm{\nabla}\times\bm{M}_{b}=\frac{e^{2}}{16\pi m^{*}}\hat{z}\times\bm{\nabla}\delta b=\Delta\bm{j}_{0}.
\end{equation}

Summing all the corrections, we find that Eq.~(\ref{eq:jcf}) is
modified by substituting $\bm{e}$ with $\bm{e}+(\hbar/2m^{\ast})\bm{\nabla}\delta b$.
After applying the self-consistent conditions Eqs.~(\ref{eq:btorho},
\ref{eq:jtoe}), we obtain
\begin{equation}
\bm{j}=-\sigma_{\mathrm{CS}}\tilde{\sigma}^{-1}\sigma_{\mathrm{CS}}\bm{E}^{\ast}+\sigma_{\mathrm{CS}}\left[\beta\bm{E}+\frac{1}{e^{2}\tilde{\chi}_{0}}\bm{\nabla}\rho\right],\label{eq:jwithb}
\end{equation}
where $\tilde{\chi}_{0}=-m^{\ast}/2\pi\hbar^{2}$ is the static density
susceptibility of a free CF system at $\bm{q}=0$, and 
\begin{equation}
\bm{E}^{\ast}=\bm{E}+\frac{1}{e^{2}\tilde{\chi}_{0}}\bm{\nabla}\rho.
\end{equation}

Finally, we determine the electron current $\bm{j}^{\mathrm{e}}$.
We need to determine the polarization density $\bm{P}$ and the magnetization
density $\bm{M}$. The polarization density does not vanish due to
the magnetic correction in Eq.~(\ref{eq:xdotdb}). Applying the average
$-e\int[\mathrm{d}\bm{p}]Df$ to both sides of Eq.~(\ref{eq:xdotdb})
and noting $-e\int[\mathrm{d}\bm{p}]D\dot{\bm{x}}f=\bm{j}-\bm{j}^{\mathrm{m}}$
in the current context, we obtain
\begin{equation}
\bm{P}=-\frac{1}{\tilde{\omega}_{B}}\frac{e^{2}}{8\pi m^{\ast}}\bm{\nabla}\delta b.\label{eq:Pmag}
\end{equation}

Next, we determine the magnetization density $\bm{M}$. Each CF carries
a magnetic moment with respect to $B$: $\bm{m}_{B}=\partial L/\partial\bm{B}=(\bm{p}\times\dot{\bm{p}})/2eB^{2}-\hat{z}(b/B^{2})(p^{2}/2m^{\ast})+\bm{m}_{B}^{\prime}$,
where $\bm{m}_{B}^{\prime}$ denotes contributions from such as the
$B$-dependence of the effective mass or the zero-point energy of
a LL as discussed in the M$^{2}$RPA theory~\citep{simon_composite_1996,simon1998}.
Substituting Eq.~(\ref{eq:pdotdb}), we obtain $\bm{m}_{B}=\bm{m}_{\bm{b}}+\bm{m}_{B}^{\prime}$.
The terms related to driving fields in Eq.~(\ref{eq:pdotdb}) are
ignored because their contributions are of higher orders of $ql_{B}$.

The total electron current can then be obtained by applying Eq.~(\ref{eq:jeexpansion}).
We note that $\bm{j}^{\mathrm{m}}$ should be subtracted from $\bm{j}$
when applying the relation. It cancels part of the contribution from
$\bm{m}_{B}$. The net magnetization correction to the electron current
is thus contributed only by $\bm{M}^{\prime}\equiv\int[\mathrm{d}\bm{p}]D\bm{m}_{B}^{\prime}f$.
We have
\begin{equation}
\bm{j}^{\mathrm{e}}=\left(-\sigma_{\mathrm{CS}}\tilde{\sigma}^{-1}\sigma_{\mathrm{CS}}+\sigma_{\mathrm{CS}}\beta\right)\bm{E}^{\ast}+\bm{\nabla}\times\bm{M}^{\prime}.\label{eq:je}
\end{equation}
Applying the relation $\rho(\omega,\bm{q})\approx\rho^{\mathrm{e}}(\omega,\bm{q})=\bm{q}\cdot\bm{j}^{\mathrm{e}}(\omega,\bm{q})/\omega$,
we can rewrite the electron current as $\bm{j}^{\mathrm{e}}=\sigma\bm{E}+\bm{\nabla}\times\bm{M}^{\prime}$
with the conductivity tensor $\sigma$ determined by
\begin{equation}
\sigma^{-1}=\left(\sigma^{\ast}\right)^{-1}-\frac{\mathrm{i}q^{2}}{\omega}\frac{1}{e^{2}\tilde{\chi}_{0}}\left(\begin{array}{cc}
1 & 0\\
0 & 0
\end{array}\right),\label{eq:sigma}
\end{equation}
where $\sigma^{\ast}=-\sigma_{\mathrm{CS}}\tilde{\sigma}^{-1}\sigma_{\mathrm{CS}}+\sigma_{\mathrm{CS}}\beta$
is the conductivity tensor with respect to $\bm{E}^{\ast}$. 

\subsection{Density response function}

 From Eq.~(\ref{eq:sigma}), we can determine the density response
function $\chi$ of electrons by using the relation $\chi(\omega,\bm{q})=-\mathrm{i}q^{2}\sigma_{11}(\omega,\bm{q})/e^{2}\omega$.
We have 
\begin{equation}
\frac{1}{\chi}=\frac{1}{\tilde{\chi}_{0}}+\frac{\mathrm{i}\omega\tilde{\sigma}_{22}}{q^{2}}\frac{\left(2h/e\right)^{2}}{1+\frac{2h}{e^{2}}\frac{\mathrm{i}\omega}{\tilde{\omega}_{B}}\tilde{\sigma}_{22}},\label{eq:chi}
\end{equation}
where $\tilde{\sigma}_{22}$ is the transverse component of $\tilde{\sigma}$
shown in Eq.~(\ref{eq:sigma22}). 

It can be shown that $\chi$ is related to the magnetic susceptibility
of CFs. The first term in the right hand side of Eq.~(\ref{eq:chi}),
which is contributed by the magnetic correction, can be absorbed into
the second term as a correction to $\tilde{\sigma}_{22}$. In the
low-frequency limit $\omega\rightarrow0$, we have $\Delta\tilde{\sigma}_{22}\approx(e/2h)^{2}q^{2}/\mathrm{i}\omega\tilde{\chi}_{0}$.
Because the magnetic susceptibility of CFs is related to the transverse
component of the CF conductivity tensor by $\tilde{\chi}^{\mathrm{m}}=\mathrm{i}\omega[\tilde{\sigma}_{22}+\Delta\tilde{\sigma}_{22}]/q^{2}$~\citep{giuliani_vignale_2005},
we have
\begin{equation}
\frac{1}{\chi_{0}}=\left(\frac{2h}{e}\right)^{2}\tilde{\chi}_{0}^{\mathrm{m}},\label{eq:chiom}
\end{equation}
where $\chi_{0}$ ($\tilde{\chi}_{0}^{\mathrm{m}}$) denotes the static
density (magnetic) susceptibility of electrons (CFs).

Equation~(\ref{eq:chiom}) implies a self-consistency condition for
the model of CFs. In Sec.~\ref{subsec:CS magnetic fluctuations},
we assume the relation $\delta b(\bm{x})=-(2m^{*}/e\hbar)V_{\textrm{eff}}(\bm{x})$
for determining the magnitude of the fluctuating CS magnetic field.
It leads to the conclusion of no-side-jumps in Sec.~\ref{subsec:CS magnetic fluctuations},
and our derivation of the extrinsic response is based on the conclusion.
The relation requires that the density response function of electrons
should have the long-wavelength static limit $\chi_{0}=\tilde{\chi}_{0}$.
Therefore, to make our considerations self-consistent, the static
density susceptibility $\tilde{\chi}_{0}$ of CFs should be related
to $\tilde{\chi}_{0}^{\mathrm{m}}$ by the same relation as Eq.~(\ref{eq:chiom}).
It is easy to verify that the relation indeed holds for our semiclassical
model. However, our semi-classical approach cannot fully determine
the magnetic susceptibility without taking account of quantum effects
(e.g., Landau diamagnetism) and specifying an effective hamiltonian
to the quadratic order of $\delta b$~\citep{gao2015}. Therefore,
we need to assume that a full effective hamiltonian of CFs has these
corrections canceled.

\section{An experimental test\label{sec:An-experimental-test}}

The longitudinal conductivity $\sigma_{11}(\omega,\bm{q})$ of a two
dimensional electron gas (2DEG) at finite $\bm{q}$ can be measured
by using a surface acoustic wave (SAW) that traverses the 2DEG. The
SAW is attenuated by the 2DEG and its velocity is altered~\citep{wixforth1989a,willett1990,willett_enhanced_1993}.
It is shown that the shift of the sound velocity $v_{\mathrm{s}}$
is related to $\sigma_{11}(\omega,\bm{q})$ by~\citep{ridley1988,wixforth1989a,mirlin_composite_1997}
\begin{equation}
\frac{\Delta v_{\mathrm{s}}}{v_{\mathrm{s}}}=\frac{\alpha^{2}}{2}\mathrm{Re}\frac{1}{1+\mathrm{i}\sigma_{11}(\omega,\bm{q})/\sigma_{\mathrm{m}}},\label{eq:rdv}
\end{equation}
where $\alpha^{2}/2=3.2\times10^{-4}$ is the piezoelectric coupling
constant for GaAs, $\sigma_{\mathrm{m}}=2\epsilon_{\mathrm{eff}}v_{\mathrm{s}}$,
and $\epsilon_{\mathrm{eff}}$ is the effective dielectric constant
for the 2DEG.

The dipole theory (as well as the Dirac CF theory) predicts a longitudinal
conductivity $\sigma_{11}(\omega,\bm{q})$ different from that of
the HLR theory. The most important difference is contributed by the
off-diagonal component of the $\beta$ matrix in Eq.~(\ref{eq:DiracSigmaTrans})
which originates from the dipole correction (see Sec.~\ref{sec:CF-current}).
We have:
\begin{equation}
\sigma_{11}^{\mathrm{Dipole}}-\sigma_{11}^{\mathrm{HLR}}\approx\mathrm{i}\frac{e^{2}}{2h}\frac{\omega}{\tilde{\omega}_{\mathrm{B}}}.\label{eq:dpcorrection}
\end{equation}
The magnitude of the correction ($\sim3\times10^{-7}\mathrm{S}$ for
the experiment considered below) turns out to be comparable to that
of $\sigma_{\mathrm{m}}$. Since the correction to the imaginary part
of the conductivity scales the velocity shift: $\Delta v_{\mathrm{s}}/v_{\mathrm{s}}\approx(\alpha^{2}/2)\mathrm{[1-\mathrm{Im}(\sigma_{11}/\sigma_{\mathrm{m}})]/[\mathrm{Re}(\sigma_{11}/\sigma_{\mathrm{m}})]^{2}}$,
there is a good chance to detect its effect in a SAW experiment.

\begin{figure}
\includegraphics[width=1\columnwidth]{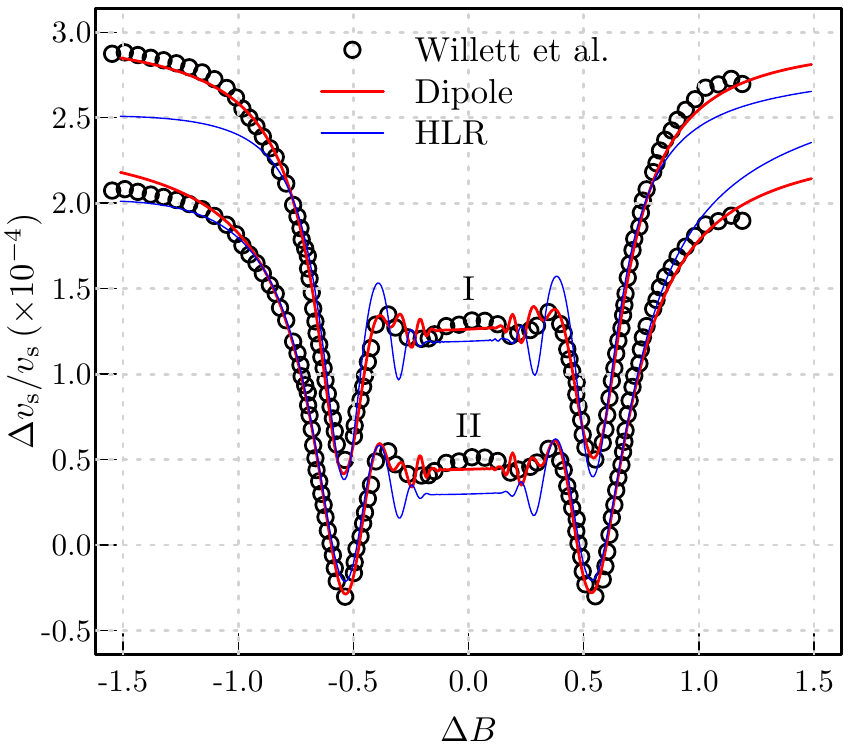}

\caption{\label{fig:fit} Experimental data of the SAW velocity shift adapted
from Ref.~\citep{willett1995} (black circles) and fits to the data
by using the dipole theory (red thick lines) and the HLR theory (blue
thin lines). Two different ways of fittings are shown: (I) fits the
data as it is by treating $\sigma_{\mathrm{m}}$ as a fitting parameter;
(II) fits the data after subtracting a constant background and fixes
the value of $\sigma_{\mathrm{m}}$ to $6.8\times10^{-4}\mathrm{S}$.
For clarity, (II) is offset vertically by $-0.8\times10^{-4}$.}
\end{figure}

As a test of our theory and an effort of looking for the dipole correction,
we fit the data presented in Ref.~\citep{willett1995}. The experiment
is one of a series of SAW experiments carried out by Willett et~al.
on CF systems~\citep{willett_experimental_1993,willett_enhanced_1993,willett1995,willett_geometric_1999}.
It employs a high SAW frequency (10.7GHz) and thus has the best data
quality. The data is shown in Fig.~\ref{fig:fit}.

\begin{table}[t]
\begin{tabular}{cccccccc}
\toprule 
\multirow{2}{*}{} & \multirow{2}{*}{} & $v_{\mathrm{s}}$ & $m^{\ast}$ & $\tau$ & $\sigma_{\mathrm{m}}$ & $\alpha^{2}/2$ & $(\Delta v_{\mathrm{s}}/v_{\mathrm{s}})_{0}$\tabularnewline
 &  & ($10^{3}\mathrm{m/s}$) & ($m_{\mathrm{e}}$) & (ps) & ($10^{-7}\mathrm{S}$) & ($10^{-4}$) & ($10^{-4}$)\tabularnewline
\midrule
\midrule 
\multirow{2}{*}{I} & Dipole & $3.1$ & $0.6$ & $14$ & $14$ & $\underline{3.2}$ & $\underline{0}$\tabularnewline
\cmidrule{2-8} \cmidrule{3-8} \cmidrule{4-8} \cmidrule{5-8} \cmidrule{6-8} \cmidrule{7-8} \cmidrule{8-8} 
 & HLR & $3.0$ & $\underline{0.3}$ & $7.3$ & $12$ & $\underline{3.2}$ & $\underline{0}$\tabularnewline
\midrule 
\multirow{2}{*}{II} & Dipole & $3.1$ & $0.7$ & $20$ & $\underline{6.8}$ & $2.8$ & $0.89$\tabularnewline
\cmidrule{2-8} \cmidrule{3-8} \cmidrule{4-8} \cmidrule{5-8} \cmidrule{6-8} \cmidrule{7-8} \cmidrule{8-8} 
 & HLR & $3.1$ & $\underline{0.3}$ & $5.7$ & $\underline{6.8}$ & $4.4$ & $0.30$\tabularnewline
\midrule 
\multicolumn{2}{c}{Reference} & $3.0$~\citep{simon1996a} & $0.8$~\citep{willett1995} & -- & $6.8$ & $3.2$ & --\tabularnewline
\bottomrule
\end{tabular}

\caption{\label{tab:parameters} The values of fitting parameters. Underlined
numbers indicate values fixed in fittings. $(\Delta v_{\mathrm{s}}/v_{\mathrm{s}})_{0}$
is the value of the background subtracted from the data. The last
line shows the reference values of the parameters as well as their
sources.}
\end{table}

We first fit the data as it is. We find that the dipole theory fits
the data rather well. It reproduces the main features of the resonances
and coincides well with the data in the whole range of $\Delta B$,
as evident in Fig.~\ref{fig:fit} (denoted as I). The parameters
for the fit are shown in Table \ref{tab:parameters}. All parameters
except $\sigma_{\mathrm{m}}$ are close to their respective reference
values. $\sigma_{\mathrm{m}}$, on the other hand, is approximately
a factor of two larger than the expected value. In comparison, the
HLR theory fits the data less well. It cannot fit the data near $\Delta B\sim0$
and at the two ends simultaneously, and produces sub-resonance features
that are too pronounced. Moreover, we cannot treat $m^{\ast}$ as
a fitting parameter for the fitting to the HLR theory because it converges
to an unphysical negative value. We therefore set $m^{\ast}=0.3m_{\mathrm{e}}$.
Smaller values of $m^{\ast}$ yield slightly better fits but not much
different from the one shown. The fitted value of $\sigma_{\mathrm{m}}$
is also significantly larger than its expected value.

The disparity between the fitted values of $\sigma_{\mathrm{m}}$
and its expected value is an issue. In literatures, the value of $\sigma_{\mathrm{m}}$
is usually quoted as $3.5\times10^{-7}\mathrm{S}$~\citep{willett_experimental_1993,willett_enhanced_1993,willett1995,willett_geometric_1999}.
The value is obtained by assuming $\epsilon_{\mathrm{eff}}=(\epsilon_{0}+\epsilon)/2$,
where $\epsilon_{0}$ and $\epsilon$ are the dielectric constants
of vacuum and GaAs, respectively. The assumption is appropriate only
when the 2DEG is on top of a sample surface, while in reality 2DEG
is always a finite depth $d$ underneath the surface. When $qd\gtrsim1$,
which is likely true for the current case, one should use $\epsilon_{\mathrm{eff}}\approx\epsilon$~\citep{simon1996a}.
It yields the reference value $\sigma_{\mathrm{m}}=6.8\times\times10^{-7}\mathrm{S}$
shown in Table \ref{tab:parameters}. Unfortunately, it is not easy
to account for the remaining disparity in the value of $\sigma_{\mathrm{m}}$.
We are not aware of another mechanism that could increase $\sigma_{\mathrm{m}}$.

We therefore try to fit the data by insisting $\sigma_{\mathrm{m}}=6.8\times\times10^{-7}\Omega^{-1}$.
In this case, both the theories cannot fit the data as they show sizable
deviations and too strong resonances. Actually, in the regime $ql\gg1$
($ql\approx8$ in the current case) necessary for the observation
of strong resonances, we have
\begin{equation}
\sigma_{11}\left(\Delta B=0\right)\approx\frac{e^{2}}{4h}\left(\frac{q}{k_{\mathrm{F}}}\pm\mathrm{i}\frac{\omega}{\tilde{\omega}_{\mathrm{B}}}\right),
\end{equation}
where the $+$ and $-$ signs are for the dipole theory and the HLR
theory, respectively. The real (and dominant) part of the conductivity
is set only by external parameters (the wavenumber and the density).
It predicts $\Delta v_{\mathrm{s}}/v_{\mathrm{s}}\sim0.4\times10^{-4}$
at $\Delta B=0$ for $\sigma_{\mathrm{m}}=6.8\times10^{-7}\mathrm{S}$,
much lower than the value actually observed in the experiment. Moreover,
because $\sigma_{11}$ is determined by the transverse component of
the CF conductivity $\tilde{\sigma}_{22}$ whose real part is not
renormalized by interaction in the clean limit~\citep{nozieres1997},
the disparity cannot be eliminated by considering a residual interaction
between CFs. 

We thus conjecture that there may exist extrinsic mechanisms which
contribute a background to the experimentally measured SAW velocity
shifts. We treat the value of the background as a fitting parameter.
We also treat the piezoelectric coupling constant $\alpha^{2}/2$
as a fitting parameter because it is shown to be a function of the
depth $d$ of the 2DEG layer~\citep{simon1996a}. The fittings are
shown in Fig.~\ref{fig:fit} with parameters shown in Table~\ref{tab:parameters}
(denoted as II). It is evident that the dipole theory has the better
fit. It yields parameters close to their respective reference values
or in the reasonable regime. On the contrary, in the fitting to the
HLR theory, we find sizable deviations and a fitted value of $\alpha^{2}/2$
which is too large~\footnote{The constant has an oscillatory dependence on $d$. Our own calculation
shows the same qualitative behavior as that shown in Ref~\citep{simon1996a}
but with quantitative differences. It indicates that the constant
has a maximal value $3.7\times10^{-4}$ at $d=0$ and a second peak
$3.1\times10^{-4}$ at $qd\sim2.5$.}.

Although the dipole theory consistently provides better fits to the
experimental data, the support to it cannot be regarded conclusive
before either the value of $\sigma_{\mathrm{m}}$ or the conjectured
background can be clarified. We also apply the fitting to the set
of data presented in Ref.~\citep{willett_experimental_1993}. We
find that both the theories can reproduce main features but with sizable
quantitative deviations. More experiments may be needed for clarifying
various factors that may affect the measurement. Our analysis, nevertheless,
does show that SAW experiments have the potential for measuring the
dipole correction and differentiating different theories of CFs.

Finally, we note that the magnetic correction shown in Eq.~(\ref{eq:sigma})
only slightly modifies $\sigma_{11}$, and the quadrupole correction,
which only alters the Hall conductivity, does not have an effect to
the velocity shift. Therefore, SAW experiments can not distinguish
the dipole theory from the Dirac CF theory.

\section{Summary and discussion\label{sec:concluding-remarks}}

In summary, we determine the EM response of CFs of the dipole picture.
We show how the Dirac CF-like EM response emerges in the dipole picture
(Sec.~\ref{subsec:Emergence-of-the}). We show that a CF system has
an intrinsic Hall response which is independent of the filling factor
(Sec.~\ref{subsec:Intrinsic-Hall-conductance}) and not altered by
impurity scattering (Sec.~\ref{sec:Scattering of CF}). When one
ascribes the intrinsic response to a redefined vacuum, the remainder
of the EM response can be interpreted as from a Dirac CF system as
far as the the low energy and long wavelength response is concerned.

When we go beyond the zeroth order in $ql_{B}$, deviations from the
Dirac CF theory start to show up (Sec.~\ref{sec:Quadrupole-and-magnetic}).
We see that the quadrupole correction gives rise to a correction to
the Hall conductivity, and the magnetic correction modifies the density
response function. Although it could be hard to detect these corrections
in experiments, they do manifest an important fact of the dipole picture,
i.e., a CF has an internal structure.

It is natural to ask whether or not the internal structure of a CF
can be probed in experiments. In the dipole picture, a CF has a length
scale $l_{B}$, comparable to its Fermi wavelength. It suggests that
carefully designed experiments that can measure the Fermi wavelength
could potentially probe the internal structure as well. To this end,
we fit the SAW measurement data by Willett et~al., and show that
the contribution from the dipole correction is sizable and improves
fittings. To distinguish the dipole model and the Dirac CF theory,
on the other hand, we need to detect quadrupoles or higher order corrections.
In Ref.~\citep{ji_asymmetry_2020}, we propose such an experiment.
We show that a geometrical resonance experiment with a modulated magnetic
field will give rise to asymmetry opposite to that observed in a similar
experiment but with a modulated scalar potential (or the CS magnetic
field).

Our theory suggests a couple of constraints on the effective model
describing the dipole, including the density-of-state factor appeared
in the dispersion Eq.~(\ref{eq:dispersion}) and the relation Eq.~(\ref{eq:chiom})
between the magnetic and density susceptibilities. In principle, one
could infer an effective Hamiltonian directly from microscopic wave-functions.
It is necessary to develop a scheme to do that and test the viability
of our assumptions. This will be left as a future investigation.

We do not consider the effects of residual interaction between CFs
for a reason. For both the HLR theory and the Dirac CF theory, the
consideration is essential because both the theories rely on renormalizations
to eliminate the unwelcome presences of bare parameters, i.e., the
band mass for the HLR theory and the velocity of the massless Dirac
cone for the Dirac CF theory. For our theory, on the other hand, the
consideration is less important because our model is inferred directly
from a microscopic CF wave function defined in the projected Hilbert
space of a LL. Spurious dependences on bare parameters have been eliminated
from the beginning. The residual interaction between CFs, which should
also be derived directly from the microscopic CF wave function, would
renormalize model parameters. In contrast to the HLR theory and the
Dirac CF theory, the renormalization in our theory needs not relate
the parameters to their bare counterparts.
\begin{acknowledgments}
This work is supported by the National Basic Research Program of China
(973 Program) Grants No. 2018YFA0305603 and No. 2015CB921101 and the
National Science Foundation of China Grant No. 11325416.
\end{acknowledgments}

\bibliographystyle{apsrev4-1}
\bibliography{composite_fermion,CompositeFermions}

\begin{thebibliography}{60}%
\makeatletter
\providecommand \@ifxundefined [1]{%
 \@ifx{#1\undefined}
}%
\providecommand \@ifnum [1]{%
 \ifnum #1\expandafter \@firstoftwo
 \else \expandafter \@secondoftwo
 \fi
}%
\providecommand \@ifx [1]{%
 \ifx #1\expandafter \@firstoftwo
 \else \expandafter \@secondoftwo
 \fi
}%
\providecommand \natexlab [1]{#1}%
\providecommand \enquote  [1]{``#1''}%
\providecommand \bibnamefont  [1]{#1}%
\providecommand \bibfnamefont [1]{#1}%
\providecommand \citenamefont [1]{#1}%
\providecommand \href@noop [0]{\@secondoftwo}%
\providecommand \href [0]{\begingroup \@sanitize@url \@href}%
\providecommand \@href[1]{\@@startlink{#1}\@@href}%
\providecommand \@@href[1]{\endgroup#1\@@endlink}%
\providecommand \@sanitize@url [0]{\catcode `\\12\catcode `\$12\catcode
  `\&12\catcode `\#12\catcode `\^12\catcode `\_12\catcode `\%12\relax}%
\providecommand \@@startlink[1]{}%
\providecommand \@@endlink[0]{}%
\providecommand \url  [0]{\begingroup\@sanitize@url \@url }%
\providecommand \@url [1]{\endgroup\@href {#1}{\urlprefix }}%
\providecommand \urlprefix  [0]{URL }%
\providecommand \Eprint [0]{\href }%
\providecommand \doibase [0]{http://dx.doi.org/}%
\providecommand \selectlanguage [0]{\@gobble}%
\providecommand \bibinfo  [0]{\@secondoftwo}%
\providecommand \bibfield  [0]{\@secondoftwo}%
\providecommand \translation [1]{[#1]}%
\providecommand \BibitemOpen [0]{}%
\providecommand \bibitemStop [0]{}%
\providecommand \bibitemNoStop [0]{.\EOS\space}%
\providecommand \EOS [0]{\spacefactor3000\relax}%
\providecommand \BibitemShut  [1]{\csname bibitem#1\endcsname}%
\let\auto@bib@innerbib\@empty
\bibitem [{\citenamefont {Tsui}\ \emph {et~al.}(1982)\citenamefont {Tsui},
  \citenamefont {Stormer},\ and\ \citenamefont
  {Gossard}}]{tsui_two-dimensional_1982}%
  \BibitemOpen
  \bibfield  {author} {\bibinfo {author} {\bibfnamefont {D.~C.}\ \bibnamefont
  {Tsui}}, \bibinfo {author} {\bibfnamefont {H.~L.}\ \bibnamefont {Stormer}}, \
  and\ \bibinfo {author} {\bibfnamefont {A.~C.}\ \bibnamefont {Gossard}},\
  }\href {\doibase 10.1103/PhysRevLett.48.1559} {\bibfield  {journal} {\bibinfo
   {journal} {Phys. Rev. Lett.}\ }\textbf {\bibinfo {volume} {48}},\ \bibinfo
  {pages} {1559} (\bibinfo {year} {1982})}\BibitemShut {NoStop}%
\bibitem [{\citenamefont {Jain}(2007)}]{jain2007composite}%
  \BibitemOpen
  \bibfield  {author} {\bibinfo {author} {\bibfnamefont {J.}~\bibnamefont
  {Jain}},\ }\href {https://books.google.com.hk/books?id=0jv9UF6UL20C} {\emph
  {\bibinfo {title} {Composite Fermions}}}\ (\bibinfo  {publisher} {Cambridge
  University Press},\ \bibinfo {year} {2007})\BibitemShut {NoStop}%
\bibitem [{\citenamefont {Jain}\ and\ \citenamefont
  {Anderson}(2009)}]{jain_beyond_2009}%
  \BibitemOpen
  \bibfield  {author} {\bibinfo {author} {\bibfnamefont {J.~K.}\ \bibnamefont
  {Jain}}\ and\ \bibinfo {author} {\bibfnamefont {P.~W.}\ \bibnamefont
  {Anderson}},\ }\href {\doibase 10.1073/pnas.0902901106} {\bibfield  {journal}
  {\bibinfo  {journal} {Proceedings of the National Academy of Sciences}\
  }\textbf {\bibinfo {volume} {106}},\ \bibinfo {pages} {9131} (\bibinfo {year}
  {2009})}\BibitemShut {NoStop}%
\bibitem [{\citenamefont {Willet}(1998)}]{willet1998}%
  \BibitemOpen
  \bibfield  {author} {\bibinfo {author} {\bibfnamefont {R.}~\bibnamefont
  {Willet}},\ }in\ \href@noop {} {\emph {\bibinfo {booktitle} {Composite
  {{Fermions}}}}},\ \bibinfo {editor} {edited by\ \bibinfo {editor}
  {\bibfnamefont {O.}~\bibnamefont {Heinonen}}}\ (\bibinfo  {publisher} {{World
  Scientific}},\ \bibinfo {year} {1998})\ p.\ \bibinfo {pages}
  {349}\BibitemShut {NoStop}%
\bibitem [{\citenamefont {Smet}(1998)}]{smet1998}%
  \BibitemOpen
  \bibfield  {author} {\bibinfo {author} {\bibfnamefont {J.~H.}\ \bibnamefont
  {Smet}},\ }in\ \href@noop {} {\emph {\bibinfo {booktitle} {Composite
  {{Fermions}}}}},\ \bibinfo {editor} {edited by\ \bibinfo {editor}
  {\bibfnamefont {O.}~\bibnamefont {Heinonen}}}\ (\bibinfo  {publisher} {{World
  Scientific}},\ \bibinfo {year} {1998})\ p.\ \bibinfo {pages}
  {443}\BibitemShut {NoStop}%
\bibitem [{\citenamefont {Willett}\ \emph {et~al.}(1999)\citenamefont
  {Willett}, \citenamefont {West},\ and\ \citenamefont
  {Pfeiffer}}]{willett_geometric_1999}%
  \BibitemOpen
  \bibfield  {author} {\bibinfo {author} {\bibfnamefont {R.~L.}\ \bibnamefont
  {Willett}}, \bibinfo {author} {\bibfnamefont {K.~W.}\ \bibnamefont {West}}, \
  and\ \bibinfo {author} {\bibfnamefont {L.~N.}\ \bibnamefont {Pfeiffer}},\
  }\href {\doibase 10.1103/PhysRevLett.83.2624} {\bibfield  {journal} {\bibinfo
   {journal} {Phys. Rev. Lett.}\ }\textbf {\bibinfo {volume} {83}},\ \bibinfo
  {pages} {2624} (\bibinfo {year} {1999})}\BibitemShut {NoStop}%
\bibitem [{\citenamefont {Smet}\ \emph {et~al.}(1999)\citenamefont {Smet},
  \citenamefont {Jobst}, \citenamefont {von Klitzing}, \citenamefont {Weiss},
  \citenamefont {Wegscheider},\ and\ \citenamefont
  {Umansky}}]{smet_commensurate_1999}%
  \BibitemOpen
  \bibfield  {author} {\bibinfo {author} {\bibfnamefont {J.~H.}\ \bibnamefont
  {Smet}}, \bibinfo {author} {\bibfnamefont {S.}~\bibnamefont {Jobst}},
  \bibinfo {author} {\bibfnamefont {K.}~\bibnamefont {von Klitzing}}, \bibinfo
  {author} {\bibfnamefont {D.}~\bibnamefont {Weiss}}, \bibinfo {author}
  {\bibfnamefont {W.}~\bibnamefont {Wegscheider}}, \ and\ \bibinfo {author}
  {\bibfnamefont {V.}~\bibnamefont {Umansky}},\ }\href {\doibase
  10.1103/PhysRevLett.83.2620} {\bibfield  {journal} {\bibinfo  {journal}
  {Phys. Rev. Lett.}\ }\textbf {\bibinfo {volume} {83}},\ \bibinfo {pages}
  {2620} (\bibinfo {year} {1999})}\BibitemShut {NoStop}%
\bibitem [{\citenamefont {Halperin}\ \emph {et~al.}(1993)\citenamefont
  {Halperin}, \citenamefont {Lee},\ and\ \citenamefont
  {Read}}]{halperin_theory_1993}%
  \BibitemOpen
  \bibfield  {author} {\bibinfo {author} {\bibfnamefont {B.~I.}\ \bibnamefont
  {Halperin}}, \bibinfo {author} {\bibfnamefont {P.~A.}\ \bibnamefont {Lee}}, \
  and\ \bibinfo {author} {\bibfnamefont {N.}~\bibnamefont {Read}},\ }\href
  {\doibase 10.1103/PhysRevB.47.7312} {\bibfield  {journal} {\bibinfo
  {journal} {Phys. Rev. B}\ }\textbf {\bibinfo {volume} {47}},\ \bibinfo
  {pages} {7312} (\bibinfo {year} {1993})}\BibitemShut {NoStop}%
\bibitem [{\citenamefont {Zhang}\ \emph {et~al.}(1989)\citenamefont {Zhang},
  \citenamefont {Hansson},\ and\ \citenamefont
  {Kivelson}}]{zhang_effective-field-theory_1989}%
  \BibitemOpen
  \bibfield  {author} {\bibinfo {author} {\bibfnamefont {S.~C.}\ \bibnamefont
  {Zhang}}, \bibinfo {author} {\bibfnamefont {T.~H.}\ \bibnamefont {Hansson}},
  \ and\ \bibinfo {author} {\bibfnamefont {S.}~\bibnamefont {Kivelson}},\
  }\href {\doibase 10.1103/PhysRevLett.62.82} {\bibfield  {journal} {\bibinfo
  {journal} {Phys. Rev. Lett.}\ }\textbf {\bibinfo {volume} {62}},\ \bibinfo
  {pages} {82} (\bibinfo {year} {1989})}\BibitemShut {NoStop}%
\bibitem [{\citenamefont {Lopez}\ and\ \citenamefont
  {Fradkin}(1991)}]{lopez_fractional_1991}%
  \BibitemOpen
  \bibfield  {author} {\bibinfo {author} {\bibfnamefont {A.}~\bibnamefont
  {Lopez}}\ and\ \bibinfo {author} {\bibfnamefont {E.}~\bibnamefont
  {Fradkin}},\ }\href {\doibase 10.1103/PhysRevB.44.5246} {\bibfield  {journal}
  {\bibinfo  {journal} {Phys. Rev. B}\ }\textbf {\bibinfo {volume} {44}},\
  \bibinfo {pages} {5246} (\bibinfo {year} {1991})}\BibitemShut {NoStop}%
\bibitem [{\citenamefont {Kalmeyer}\ and\ \citenamefont
  {Zhang}(1992)}]{kalmeyer_metallic_1992}%
  \BibitemOpen
  \bibfield  {author} {\bibinfo {author} {\bibfnamefont {V.}~\bibnamefont
  {Kalmeyer}}\ and\ \bibinfo {author} {\bibfnamefont {S.-C.}\ \bibnamefont
  {Zhang}},\ }\href {\doibase 10.1103/PhysRevB.46.9889} {\bibfield  {journal}
  {\bibinfo  {journal} {Phys. Rev. B}\ }\textbf {\bibinfo {volume} {46}},\
  \bibinfo {pages} {9889} (\bibinfo {year} {1992})}\BibitemShut {NoStop}%
\bibitem [{\citenamefont {Simon}\ and\ \citenamefont
  {Halperin}(1993)}]{simon_finite-wave-vector_1993}%
  \BibitemOpen
  \bibfield  {author} {\bibinfo {author} {\bibfnamefont {S.~H.}\ \bibnamefont
  {Simon}}\ and\ \bibinfo {author} {\bibfnamefont {B.~I.}\ \bibnamefont
  {Halperin}},\ }\href {\doibase 10.1103/PhysRevB.48.17368} {\bibfield
  {journal} {\bibinfo  {journal} {Phys. Rev. B}\ }\textbf {\bibinfo {volume}
  {48}},\ \bibinfo {pages} {17368} (\bibinfo {year} {1993})}\BibitemShut
  {NoStop}%
\bibitem [{\citenamefont {Simon}(1998)}]{simon1998}%
  \BibitemOpen
  \bibfield  {author} {\bibinfo {author} {\bibfnamefont {S.~H.}\ \bibnamefont
  {Simon}},\ }in\ \href@noop {} {\emph {\bibinfo {booktitle} {Composite
  {{Fermions}}}}},\ \bibinfo {editor} {edited by\ \bibinfo {editor}
  {\bibfnamefont {O.}~\bibnamefont {Heinonen}}}\ (\bibinfo  {publisher} {{World
  Scientific}},\ \bibinfo {year} {1998})\ p.~\bibinfo {pages} {91}\BibitemShut
  {NoStop}%
\bibitem [{\citenamefont {Kivelson}\ \emph {et~al.}(1997)\citenamefont
  {Kivelson}, \citenamefont {Lee}, \citenamefont {Krotov},\ and\ \citenamefont
  {Gan}}]{kivelson_composite-fermion_1997}%
  \BibitemOpen
  \bibfield  {author} {\bibinfo {author} {\bibfnamefont {S.~A.}\ \bibnamefont
  {Kivelson}}, \bibinfo {author} {\bibfnamefont {D.-H.}\ \bibnamefont {Lee}},
  \bibinfo {author} {\bibfnamefont {Y.}~\bibnamefont {Krotov}}, \ and\ \bibinfo
  {author} {\bibfnamefont {J.}~\bibnamefont {Gan}},\ }\href {\doibase
  10.1103/PhysRevB.55.15552} {\bibfield  {journal} {\bibinfo  {journal} {Phys.
  Rev. B}\ }\textbf {\bibinfo {volume} {55}},\ \bibinfo {pages} {15552}
  (\bibinfo {year} {1997})}\BibitemShut {NoStop}%
\bibitem [{\citenamefont {Wang}\ \emph {et~al.}(2017)\citenamefont {Wang},
  \citenamefont {Cooper}, \citenamefont {Halperin},\ and\ \citenamefont
  {Stern}}]{wang_particle-hole_2017}%
  \BibitemOpen
  \bibfield  {author} {\bibinfo {author} {\bibfnamefont {C.}~\bibnamefont
  {Wang}}, \bibinfo {author} {\bibfnamefont {N.~R.}\ \bibnamefont {Cooper}},
  \bibinfo {author} {\bibfnamefont {B.~I.}\ \bibnamefont {Halperin}}, \ and\
  \bibinfo {author} {\bibfnamefont {A.}~\bibnamefont {Stern}},\ }\href
  {\doibase 10.1103/PhysRevX.7.031029} {\bibfield  {journal} {\bibinfo
  {journal} {Phys. Rev. X}\ }\textbf {\bibinfo {volume} {7}},\ \bibinfo {pages}
  {031029} (\bibinfo {year} {2017})}\BibitemShut {NoStop}%
\bibitem [{\citenamefont {Son}(2015)}]{son_is_2015}%
  \BibitemOpen
  \bibfield  {author} {\bibinfo {author} {\bibfnamefont {D.~T.}\ \bibnamefont
  {Son}},\ }\href {\doibase 10.1103/PhysRevX.5.031027} {\bibfield  {journal}
  {\bibinfo  {journal} {Phys. Rev. X}\ }\textbf {\bibinfo {volume} {5}},\
  \bibinfo {pages} {031027} (\bibinfo {year} {2015})}\BibitemShut {NoStop}%
\bibitem [{\citenamefont {Geraedts}\ \emph {et~al.}(2016)\citenamefont
  {Geraedts}, \citenamefont {Zaletel}, \citenamefont {Mong}, \citenamefont
  {Metlitski}, \citenamefont {Vishwanath},\ and\ \citenamefont
  {Motrunich}}]{geraedts_half-filled_2016}%
  \BibitemOpen
  \bibfield  {author} {\bibinfo {author} {\bibfnamefont {S.~D.}\ \bibnamefont
  {Geraedts}}, \bibinfo {author} {\bibfnamefont {M.~P.}\ \bibnamefont
  {Zaletel}}, \bibinfo {author} {\bibfnamefont {R.~S.~K.}\ \bibnamefont
  {Mong}}, \bibinfo {author} {\bibfnamefont {M.~A.}\ \bibnamefont {Metlitski}},
  \bibinfo {author} {\bibfnamefont {A.}~\bibnamefont {Vishwanath}}, \ and\
  \bibinfo {author} {\bibfnamefont {O.~I.}\ \bibnamefont {Motrunich}},\ }\href
  {\doibase 10.1126/science.aad4302} {\bibfield  {journal} {\bibinfo  {journal}
  {Science}\ }\textbf {\bibinfo {volume} {352}},\ \bibinfo {pages} {197}
  (\bibinfo {year} {2016})}\BibitemShut {NoStop}%
\bibitem [{\citenamefont {Pan}\ \emph {et~al.}(2017)\citenamefont {Pan},
  \citenamefont {Kang}, \citenamefont {Baldwin}, \citenamefont {West},
  \citenamefont {Pfeiffer},\ and\ \citenamefont {Tsui}}]{pan_berry_2017}%
  \BibitemOpen
  \bibfield  {author} {\bibinfo {author} {\bibfnamefont {W.}~\bibnamefont
  {Pan}}, \bibinfo {author} {\bibfnamefont {W.}~\bibnamefont {Kang}}, \bibinfo
  {author} {\bibfnamefont {K.~W.}\ \bibnamefont {Baldwin}}, \bibinfo {author}
  {\bibfnamefont {K.~W.}\ \bibnamefont {West}}, \bibinfo {author}
  {\bibfnamefont {L.~N.}\ \bibnamefont {Pfeiffer}}, \ and\ \bibinfo {author}
  {\bibfnamefont {D.~C.}\ \bibnamefont {Tsui}},\ }\href {\doibase
  10.1038/nphys4231} {\bibfield  {journal} {\bibinfo  {journal} {Nature
  Physics}\ }\textbf {\bibinfo {volume} {13}},\ \bibinfo {pages} {1168}
  (\bibinfo {year} {2017})}\BibitemShut {NoStop}%
\bibitem [{\citenamefont {Mross}\ \emph {et~al.}(2016)\citenamefont {Mross},
  \citenamefont {Alicea},\ and\ \citenamefont {Motrunich}}]{mross2016}%
  \BibitemOpen
  \bibfield  {author} {\bibinfo {author} {\bibfnamefont {D.~F.}\ \bibnamefont
  {Mross}}, \bibinfo {author} {\bibfnamefont {J.}~\bibnamefont {Alicea}}, \
  and\ \bibinfo {author} {\bibfnamefont {O.~I.}\ \bibnamefont {Motrunich}},\
  }\href {\doibase 10.1103/PhysRevLett.117.016802} {\bibfield  {journal}
  {\bibinfo  {journal} {Phys. Rev. Lett.}\ }\textbf {\bibinfo {volume} {117}},\
  \bibinfo {pages} {016802} (\bibinfo {year} {2016})}\BibitemShut {NoStop}%
\bibitem [{\citenamefont {Levin}\ and\ \citenamefont {Son}(2017)}]{levin2017}%
  \BibitemOpen
  \bibfield  {author} {\bibinfo {author} {\bibfnamefont {M.}~\bibnamefont
  {Levin}}\ and\ \bibinfo {author} {\bibfnamefont {D.~T.}\ \bibnamefont
  {Son}},\ }\href {\doibase 10.1103/PhysRevB.95.125120} {\bibfield  {journal}
  {\bibinfo  {journal} {Phys. Rev. B}\ }\textbf {\bibinfo {volume} {95}},\
  \bibinfo {pages} {125120} (\bibinfo {year} {2017})}\BibitemShut {NoStop}%
\bibitem [{\citenamefont {Read}(1994)}]{read_theory_1994}%
  \BibitemOpen
  \bibfield  {author} {\bibinfo {author} {\bibfnamefont {N.}~\bibnamefont
  {Read}},\ }\href {\doibase 10.1088/0268-1242/9/11S/002} {\bibfield  {journal}
  {\bibinfo  {journal} {Semicond. Sci. Technol.}\ }\textbf {\bibinfo {volume}
  {9}},\ \bibinfo {pages} {1859} (\bibinfo {year} {1994})}\BibitemShut
  {NoStop}%
\bibitem [{\citenamefont {Rezayi}\ and\ \citenamefont
  {Read}(1994)}]{rezayi_fermi-liquid-like_1994}%
  \BibitemOpen
  \bibfield  {author} {\bibinfo {author} {\bibfnamefont {E.}~\bibnamefont
  {Rezayi}}\ and\ \bibinfo {author} {\bibfnamefont {N.}~\bibnamefont {Read}},\
  }\href {\doibase 10.1103/PhysRevLett.72.900} {\bibfield  {journal} {\bibinfo
  {journal} {Phys. Rev. Lett.}\ }\textbf {\bibinfo {volume} {72}},\ \bibinfo
  {pages} {900} (\bibinfo {year} {1994})}\BibitemShut {NoStop}%
\bibitem [{\citenamefont {Balram}\ and\ \citenamefont
  {Jain}(2016)}]{balram_nature_2016}%
  \BibitemOpen
  \bibfield  {author} {\bibinfo {author} {\bibfnamefont {A.~C.}\ \bibnamefont
  {Balram}}\ and\ \bibinfo {author} {\bibfnamefont {J.~K.}\ \bibnamefont
  {Jain}},\ }\href {\doibase 10.1103/PhysRevB.93.235152} {\bibfield  {journal}
  {\bibinfo  {journal} {Phys. Rev. B}\ }\textbf {\bibinfo {volume} {93}},\
  \bibinfo {pages} {235152} (\bibinfo {year} {2016})}\BibitemShut {NoStop}%
\bibitem [{\citenamefont {Shankar}\ and\ \citenamefont
  {Murthy}(1997)}]{shankar_towards_1997}%
  \BibitemOpen
  \bibfield  {author} {\bibinfo {author} {\bibfnamefont {R.}~\bibnamefont
  {Shankar}}\ and\ \bibinfo {author} {\bibfnamefont {G.}~\bibnamefont
  {Murthy}},\ }\href {\doibase 10.1103/PhysRevLett.79.4437} {\bibfield
  {journal} {\bibinfo  {journal} {Phys. Rev. Lett.}\ }\textbf {\bibinfo
  {volume} {79}},\ \bibinfo {pages} {4437} (\bibinfo {year}
  {1997})}\BibitemShut {NoStop}%
\bibitem [{\citenamefont {Shankar}(1999)}]{shankar_hamiltonian_1999}%
  \BibitemOpen
  \bibfield  {author} {\bibinfo {author} {\bibfnamefont {R.}~\bibnamefont
  {Shankar}},\ }\href {\doibase 10.1103/PhysRevLett.83.2382} {\bibfield
  {journal} {\bibinfo  {journal} {Phys. Rev. Lett.}\ }\textbf {\bibinfo
  {volume} {83}},\ \bibinfo {pages} {2382} (\bibinfo {year}
  {1999})}\BibitemShut {NoStop}%
\bibitem [{\citenamefont {Murthy}\ and\ \citenamefont
  {Shankar}(2003)}]{murthy_hamiltonian_2003}%
  \BibitemOpen
  \bibfield  {author} {\bibinfo {author} {\bibfnamefont {G.}~\bibnamefont
  {Murthy}}\ and\ \bibinfo {author} {\bibfnamefont {R.}~\bibnamefont
  {Shankar}},\ }\href {\doibase 10.1103/RevModPhys.75.1101} {\bibfield
  {journal} {\bibinfo  {journal} {Rev. Mod. Phys.}\ }\textbf {\bibinfo {volume}
  {75}},\ \bibinfo {pages} {1101} (\bibinfo {year} {2003})}\BibitemShut
  {NoStop}%
\bibitem [{\citenamefont {Gočanin}\ \emph {et~al.}(2021)\citenamefont
  {Gočanin}, \citenamefont {Predin}, \citenamefont {Ćirić}, \citenamefont
  {Radovanović},\ and\ \citenamefont
  {Milovanović}}]{gocanin_microscopic_2021}%
  \BibitemOpen
  \bibfield  {author} {\bibinfo {author} {\bibfnamefont {D.}~\bibnamefont
  {Gočanin}}, \bibinfo {author} {\bibfnamefont {S.}~\bibnamefont {Predin}},
  \bibinfo {author} {\bibfnamefont {M.~D.}\ \bibnamefont {Ćirić}}, \bibinfo
  {author} {\bibfnamefont {V.}~\bibnamefont {Radovanović}}, \ and\ \bibinfo
  {author} {\bibfnamefont {M.}~\bibnamefont {Milovanović}},\ }\href
  {http://arxiv.org/abs/2102.11313} {\bibfield  {journal} {\bibinfo  {journal}
  {arXiv:2102.11313}\ } (\bibinfo {year} {2021})}\BibitemShut {NoStop}%
\bibitem [{\citenamefont {Shi}\ and\ \citenamefont
  {Ji}(2018)}]{shi_dynamics_2018}%
  \BibitemOpen
  \bibfield  {author} {\bibinfo {author} {\bibfnamefont {J.}~\bibnamefont
  {Shi}}\ and\ \bibinfo {author} {\bibfnamefont {W.}~\bibnamefont {Ji}},\
  }\href {\doibase 10.1103/PhysRevB.97.125133} {\bibfield  {journal} {\bibinfo
  {journal} {Phys. Rev. B}\ }\textbf {\bibinfo {volume} {97}},\ \bibinfo
  {pages} {125133} (\bibinfo {year} {2018})}\BibitemShut {NoStop}%
\bibitem [{\citenamefont {Sundaram}\ and\ \citenamefont
  {Niu}(1999)}]{sundaram_wave-packet_1999}%
  \BibitemOpen
  \bibfield  {author} {\bibinfo {author} {\bibfnamefont {G.}~\bibnamefont
  {Sundaram}}\ and\ \bibinfo {author} {\bibfnamefont {Q.}~\bibnamefont {Niu}},\
  }\href {\doibase 10.1103/PhysRevB.59.14915} {\bibfield  {journal} {\bibinfo
  {journal} {Phys. Rev. B}\ }\textbf {\bibinfo {volume} {59}},\ \bibinfo
  {pages} {14915} (\bibinfo {year} {1999})}\BibitemShut {NoStop}%
\bibitem [{\citenamefont {Xiao}\ \emph {et~al.}(2010)\citenamefont {Xiao},
  \citenamefont {Chang},\ and\ \citenamefont {Niu}}]{xiao_berry_2010}%
  \BibitemOpen
  \bibfield  {author} {\bibinfo {author} {\bibfnamefont {D.}~\bibnamefont
  {Xiao}}, \bibinfo {author} {\bibfnamefont {M.-C.}\ \bibnamefont {Chang}}, \
  and\ \bibinfo {author} {\bibfnamefont {Q.}~\bibnamefont {Niu}},\ }\href
  {\doibase 10.1103/RevModPhys.82.1959} {\bibfield  {journal} {\bibinfo
  {journal} {Rev. Mod. Phys.}\ }\textbf {\bibinfo {volume} {82}},\ \bibinfo
  {pages} {1959} (\bibinfo {year} {2010})}\BibitemShut {NoStop}%
\bibitem [{\citenamefont {Ji}\ and\ \citenamefont
  {Shi}(2020{\natexlab{a}})}]{ji_asymmetry_2020}%
  \BibitemOpen
  \bibfield  {author} {\bibinfo {author} {\bibfnamefont {G.}~\bibnamefont
  {Ji}}\ and\ \bibinfo {author} {\bibfnamefont {J.}~\bibnamefont {Shi}},\
  }\href {\doibase 10.1103/PhysRevB.101.235301} {\bibfield  {journal} {\bibinfo
   {journal} {Phys. Rev. B}\ }\textbf {\bibinfo {volume} {101}},\ \bibinfo
  {pages} {235301} (\bibinfo {year} {2020}{\natexlab{a}})}\BibitemShut
  {NoStop}%
\bibitem [{\citenamefont {Lee}(1998)}]{lee_neutral_1998}%
  \BibitemOpen
  \bibfield  {author} {\bibinfo {author} {\bibfnamefont {D.-H.}\ \bibnamefont
  {Lee}},\ }\href {\doibase 10.1103/PhysRevLett.80.4745} {\bibfield  {journal}
  {\bibinfo  {journal} {Phys. Rev. Lett.}\ }\textbf {\bibinfo {volume} {80}},\
  \bibinfo {pages} {4745} (\bibinfo {year} {1998})}\BibitemShut {NoStop}%
\bibitem [{\citenamefont {Lee}(1999)}]{lee_chern-simons_1999}%
  \BibitemOpen
  \bibfield  {author} {\bibinfo {author} {\bibfnamefont {D.-H.}\ \bibnamefont
  {Lee}},\ }\href {\doibase 10.1103/PhysRevB.60.5636} {\bibfield  {journal}
  {\bibinfo  {journal} {Phys. Rev. B}\ }\textbf {\bibinfo {volume} {60}},\
  \bibinfo {pages} {5636} (\bibinfo {year} {1999})}\BibitemShut {NoStop}%
\bibitem [{\citenamefont {Pasquier}\ and\ \citenamefont
  {Haldane}(1998)}]{pasquier_dipole_1998}%
  \BibitemOpen
  \bibfield  {author} {\bibinfo {author} {\bibfnamefont {V.}~\bibnamefont
  {Pasquier}}\ and\ \bibinfo {author} {\bibfnamefont {F.~D.~M.}\ \bibnamefont
  {Haldane}},\ }\href {\doibase 10.1016/S0550-3213(98)00069-8} {\bibfield
  {journal} {\bibinfo  {journal} {Nuclear Physics B}\ }\textbf {\bibinfo
  {volume} {516}},\ \bibinfo {pages} {719} (\bibinfo {year}
  {1998})}\BibitemShut {NoStop}%
\bibitem [{\citenamefont {von Oppen}\ \emph {et~al.}(1999)\citenamefont {von
  Oppen}, \citenamefont {Halperin}, \citenamefont {Simon},\ and\ \citenamefont
  {Stern}}]{von_oppen_half-filled_1999}%
  \BibitemOpen
  \bibfield  {author} {\bibinfo {author} {\bibfnamefont {F.}~\bibnamefont {von
  Oppen}}, \bibinfo {author} {\bibfnamefont {B.~I.}\ \bibnamefont {Halperin}},
  \bibinfo {author} {\bibfnamefont {S.~H.}\ \bibnamefont {Simon}}, \ and\
  \bibinfo {author} {\bibfnamefont {A.}~\bibnamefont {Stern}},\ }in\ \href@noop
  {} {\emph {\bibinfo {booktitle} {Advances in {Solid} {State} {Physics}
  39}}},\ \bibinfo {editor} {edited by\ \bibinfo {editor} {\bibfnamefont
  {B.}~\bibnamefont {Kramer}}}\ (\bibinfo  {publisher} {Springer Berlin
  Heidelberg},\ \bibinfo {address} {Berlin, Heidelberg},\ \bibinfo {year}
  {1999})\ pp.\ \bibinfo {pages} {203--212}\BibitemShut {NoStop}%
\bibitem [{\citenamefont {Wang}\ and\ \citenamefont
  {Senthil}(2016)}]{wang_half-filled_2016}%
  \BibitemOpen
  \bibfield  {author} {\bibinfo {author} {\bibfnamefont {C.}~\bibnamefont
  {Wang}}\ and\ \bibinfo {author} {\bibfnamefont {T.}~\bibnamefont {Senthil}},\
  }\href {\doibase 10.1103/PhysRevB.93.085110} {\bibfield  {journal} {\bibinfo
  {journal} {Phys. Rev. B}\ }\textbf {\bibinfo {volume} {93}},\ \bibinfo
  {pages} {085110} (\bibinfo {year} {2016})}\BibitemShut {NoStop}%
\bibitem [{\citenamefont {Jackson}(1999)}]{jackson1999}%
  \BibitemOpen
  \bibfield  {author} {\bibinfo {author} {\bibfnamefont {J.~D.}\ \bibnamefont
  {Jackson}},\ }\href@noop {} {{\selectlanguage {English}\emph {\bibinfo
  {title} {Classical Electrodynamics}}}}\ (\bibinfo  {publisher} {{Wiley}},\
  \bibinfo {year} {1999})\BibitemShut {NoStop}%
\bibitem [{\citenamefont {Xiao}\ \emph {et~al.}(2005)\citenamefont {Xiao},
  \citenamefont {Shi},\ and\ \citenamefont {Niu}}]{xiao_berry_2005}%
  \BibitemOpen
  \bibfield  {author} {\bibinfo {author} {\bibfnamefont {D.}~\bibnamefont
  {Xiao}}, \bibinfo {author} {\bibfnamefont {J.}~\bibnamefont {Shi}}, \ and\
  \bibinfo {author} {\bibfnamefont {Q.}~\bibnamefont {Niu}},\ }\href {\doibase
  10.1103/PhysRevLett.95.137204} {\bibfield  {journal} {\bibinfo  {journal}
  {Phys. Rev. Lett.}\ }\textbf {\bibinfo {volume} {95}},\ \bibinfo {pages}
  {137204} (\bibinfo {year} {2005})}\BibitemShut {NoStop}%
\bibitem [{\citenamefont {Horváthy}(2002)}]{horvathy_non-commutative_2002}%
  \BibitemOpen
  \bibfield  {author} {\bibinfo {author} {\bibfnamefont {P.~A.}\ \bibnamefont
  {Horváthy}},\ }\href {\doibase 10.1006/aphy.2002.6271} {\bibfield  {journal}
  {\bibinfo  {journal} {Annals of Physics}\ }\textbf {\bibinfo {volume}
  {299}},\ \bibinfo {pages} {128} (\bibinfo {year} {2002})}\BibitemShut
  {NoStop}%
\bibitem [{\citenamefont {Geraedts}\ \emph {et~al.}(2018)\citenamefont
  {Geraedts}, \citenamefont {Wang}, \citenamefont {Rezayi},\ and\ \citenamefont
  {Haldane}}]{geraedts_berry_2018}%
  \BibitemOpen
  \bibfield  {author} {\bibinfo {author} {\bibfnamefont {S.~D.}\ \bibnamefont
  {Geraedts}}, \bibinfo {author} {\bibfnamefont {J.}~\bibnamefont {Wang}},
  \bibinfo {author} {\bibfnamefont {E.}~\bibnamefont {Rezayi}}, \ and\ \bibinfo
  {author} {\bibfnamefont {F.}~\bibnamefont {Haldane}},\ }\href {\doibase
  10.1103/PhysRevLett.121.147202} {\bibfield  {journal} {\bibinfo  {journal}
  {Phys. Rev. Lett.}\ }\textbf {\bibinfo {volume} {121}},\ \bibinfo {pages}
  {147202} (\bibinfo {year} {2018})}\BibitemShut {NoStop}%
\bibitem [{\citenamefont {Sinitsyn}\ \emph {et~al.}(2006)\citenamefont
  {Sinitsyn}, \citenamefont {Niu},\ and\ \citenamefont
  {MacDonald}}]{sinitsyn_coordinate_2006}%
  \BibitemOpen
  \bibfield  {author} {\bibinfo {author} {\bibfnamefont {N.~A.}\ \bibnamefont
  {Sinitsyn}}, \bibinfo {author} {\bibfnamefont {Q.}~\bibnamefont {Niu}}, \
  and\ \bibinfo {author} {\bibfnamefont {A.~H.}\ \bibnamefont {MacDonald}},\
  }\href {\doibase 10.1103/PhysRevB.73.075318} {\bibfield  {journal} {\bibinfo
  {journal} {Phys. Rev. B}\ }\textbf {\bibinfo {volume} {73}},\ \bibinfo
  {pages} {075318} (\bibinfo {year} {2006})}\BibitemShut {NoStop}%
\bibitem [{\citenamefont {Chang}\ and\ \citenamefont
  {Niu}(2008)}]{chang_berry_2008}%
  \BibitemOpen
  \bibfield  {author} {\bibinfo {author} {\bibfnamefont {M.-C.}\ \bibnamefont
  {Chang}}\ and\ \bibinfo {author} {\bibfnamefont {Q.}~\bibnamefont {Niu}},\
  }\href {\doibase 10.1088/0953-8984/20/19/193202} {\bibfield  {journal}
  {\bibinfo  {journal} {J. Phys.: Condens. Matter}\ }\textbf {\bibinfo {volume}
  {20}},\ \bibinfo {pages} {193202} (\bibinfo {year} {2008})}\BibitemShut
  {NoStop}%
\bibitem [{\citenamefont {Ji}\ and\ \citenamefont
  {Shi}(2020{\natexlab{b}})}]{ji_berry_2020}%
  \BibitemOpen
  \bibfield  {author} {\bibinfo {author} {\bibfnamefont {G.}~\bibnamefont
  {Ji}}\ and\ \bibinfo {author} {\bibfnamefont {J.}~\bibnamefont {Shi}},\
  }\href {\doibase 10.1103/PhysRevResearch.2.033329} {\bibfield  {journal}
  {\bibinfo  {journal} {Phys. Rev. Research}\ }\textbf {\bibinfo {volume}
  {2}},\ \bibinfo {pages} {033329} (\bibinfo {year}
  {2020}{\natexlab{b}})}\BibitemShut {NoStop}%
\bibitem [{\citenamefont {Callaway}(1991)}]{callaway1991quantum}%
  \BibitemOpen
  \bibfield  {author} {\bibinfo {author} {\bibfnamefont {J.}~\bibnamefont
  {Callaway}},\ }\href {https://books.google.com.tw/books?id=HKTvAAAAMAAJ}
  {\emph {\bibinfo {title} {Quantum theory of the solid state}}}\ (\bibinfo
  {publisher} {Academic Press},\ \bibinfo {year} {1991})\BibitemShut {NoStop}%
\bibitem [{\citenamefont {Mirlin}\ and\ \citenamefont
  {Wölfle}(1997)}]{mirlin_composite_1997}%
  \BibitemOpen
  \bibfield  {author} {\bibinfo {author} {\bibfnamefont {A.~D.}\ \bibnamefont
  {Mirlin}}\ and\ \bibinfo {author} {\bibfnamefont {P.}~\bibnamefont
  {Wölfle}},\ }\href {\doibase 10.1103/PhysRevLett.78.3717} {\bibfield
  {journal} {\bibinfo  {journal} {Phys. Rev. Lett.}\ }\textbf {\bibinfo
  {volume} {78}},\ \bibinfo {pages} {3717} (\bibinfo {year}
  {1997})}\BibitemShut {NoStop}%
\bibitem [{\citenamefont {Nguyen}\ \emph {et~al.}(2018)\citenamefont {Nguyen},
  \citenamefont {Golkar}, \citenamefont {Roberts},\ and\ \citenamefont
  {Son}}]{nguyen2018}%
  \BibitemOpen
  \bibfield  {author} {\bibinfo {author} {\bibfnamefont {D.~X.}\ \bibnamefont
  {Nguyen}}, \bibinfo {author} {\bibfnamefont {S.}~\bibnamefont {Golkar}},
  \bibinfo {author} {\bibfnamefont {M.~M.}\ \bibnamefont {Roberts}}, \ and\
  \bibinfo {author} {\bibfnamefont {D.~T.}\ \bibnamefont {Son}},\ }\href
  {\doibase 10.1103/PhysRevB.97.195314} {\bibfield  {journal} {\bibinfo
  {journal} {Phys. Rev. B}\ }\textbf {\bibinfo {volume} {97}},\ \bibinfo
  {pages} {195314} (\bibinfo {year} {2018})}\BibitemShut {NoStop}%
\bibitem [{\citenamefont {Kamburov}\ \emph {et~al.}(2013)\citenamefont
  {Kamburov}, \citenamefont {Liu}, \citenamefont {Shayegan}, \citenamefont
  {Pfeiffer}, \citenamefont {West},\ and\ \citenamefont
  {Baldwin}}]{kamburov_composite_2013}%
  \BibitemOpen
  \bibfield  {author} {\bibinfo {author} {\bibfnamefont {D.}~\bibnamefont
  {Kamburov}}, \bibinfo {author} {\bibfnamefont {Y.}~\bibnamefont {Liu}},
  \bibinfo {author} {\bibfnamefont {M.}~\bibnamefont {Shayegan}}, \bibinfo
  {author} {\bibfnamefont {L.~N.}\ \bibnamefont {Pfeiffer}}, \bibinfo {author}
  {\bibfnamefont {K.~W.}\ \bibnamefont {West}}, \ and\ \bibinfo {author}
  {\bibfnamefont {K.~W.}\ \bibnamefont {Baldwin}},\ }\href {\doibase
  10.1103/PhysRevLett.110.206801} {\bibfield  {journal} {\bibinfo  {journal}
  {Phys. Rev. Lett.}\ }\textbf {\bibinfo {volume} {110}},\ \bibinfo {pages}
  {206801} (\bibinfo {year} {2013})}\BibitemShut {NoStop}%
\bibitem [{\citenamefont {Kamburov}\ \emph {et~al.}(2014)\citenamefont
  {Kamburov}, \citenamefont {Liu}, \citenamefont {Mueed}, \citenamefont
  {Shayegan}, \citenamefont {Pfeiffer}, \citenamefont {West},\ and\
  \citenamefont {Baldwin}}]{kamburov_what_2014}%
  \BibitemOpen
  \bibfield  {author} {\bibinfo {author} {\bibfnamefont {D.}~\bibnamefont
  {Kamburov}}, \bibinfo {author} {\bibfnamefont {Y.}~\bibnamefont {Liu}},
  \bibinfo {author} {\bibfnamefont {M.}~\bibnamefont {Mueed}}, \bibinfo
  {author} {\bibfnamefont {M.}~\bibnamefont {Shayegan}}, \bibinfo {author}
  {\bibfnamefont {L.}~\bibnamefont {Pfeiffer}}, \bibinfo {author}
  {\bibfnamefont {K.}~\bibnamefont {West}}, \ and\ \bibinfo {author}
  {\bibfnamefont {K.}~\bibnamefont {Baldwin}},\ }\href {\doibase
  10.1103/PhysRevLett.113.196801} {\bibfield  {journal} {\bibinfo  {journal}
  {Phys. Rev. Lett.}\ }\textbf {\bibinfo {volume} {113}},\ \bibinfo {pages}
  {196801} (\bibinfo {year} {2014})}\BibitemShut {NoStop}%
\bibitem [{\citenamefont {Simon}\ \emph {et~al.}(1996)\citenamefont {Simon},
  \citenamefont {Stern},\ and\ \citenamefont
  {Halperin}}]{simon_composite_1996}%
  \BibitemOpen
  \bibfield  {author} {\bibinfo {author} {\bibfnamefont {S.~H.}\ \bibnamefont
  {Simon}}, \bibinfo {author} {\bibfnamefont {A.}~\bibnamefont {Stern}}, \ and\
  \bibinfo {author} {\bibfnamefont {B.~I.}\ \bibnamefont {Halperin}},\ }\href
  {\doibase 10.1103/PhysRevB.54.R11114} {\bibfield  {journal} {\bibinfo
  {journal} {Phys. Rev. B}\ }\textbf {\bibinfo {volume} {54}},\ \bibinfo
  {pages} {R11114} (\bibinfo {year} {1996})}\BibitemShut {NoStop}%
\bibitem [{\citenamefont {Giuliani}\ and\ \citenamefont
  {Vignale}(2005)}]{giuliani_vignale_2005}%
  \BibitemOpen
  \bibfield  {author} {\bibinfo {author} {\bibfnamefont {G.}~\bibnamefont
  {Giuliani}}\ and\ \bibinfo {author} {\bibfnamefont {G.}~\bibnamefont
  {Vignale}},\ }\href {\doibase 10.1017/CBO9780511619915} {\emph {\bibinfo
  {title} {Quantum Theory of the Electron Liquid}}}\ (\bibinfo  {publisher}
  {Cambridge University Press},\ \bibinfo {year} {2005})\BibitemShut {NoStop}%
\bibitem [{\citenamefont {Gao}\ \emph {et~al.}(2015)\citenamefont {Gao},
  \citenamefont {Yang},\ and\ \citenamefont {Niu}}]{gao2015}%
  \BibitemOpen
  \bibfield  {author} {\bibinfo {author} {\bibfnamefont {Y.}~\bibnamefont
  {Gao}}, \bibinfo {author} {\bibfnamefont {S.~A.}\ \bibnamefont {Yang}}, \
  and\ \bibinfo {author} {\bibfnamefont {Q.}~\bibnamefont {Niu}},\ }\href
  {\doibase 10.1103/PhysRevB.91.214405} {\bibfield  {journal} {\bibinfo
  {journal} {Phys. Rev. B}\ }\textbf {\bibinfo {volume} {91}},\ \bibinfo
  {pages} {214405} (\bibinfo {year} {2015})}\BibitemShut {NoStop}%
\bibitem [{\citenamefont {Wixforth}\ \emph {et~al.}(1989)\citenamefont
  {Wixforth}, \citenamefont {Scriba}, \citenamefont {Wassermeier},
  \citenamefont {Kotthaus}, \citenamefont {Weimann},\ and\ \citenamefont
  {Schlapp}}]{wixforth1989a}%
  \BibitemOpen
  \bibfield  {author} {\bibinfo {author} {\bibfnamefont {A.}~\bibnamefont
  {Wixforth}}, \bibinfo {author} {\bibfnamefont {J.}~\bibnamefont {Scriba}},
  \bibinfo {author} {\bibfnamefont {M.}~\bibnamefont {Wassermeier}}, \bibinfo
  {author} {\bibfnamefont {J.~P.}\ \bibnamefont {Kotthaus}}, \bibinfo {author}
  {\bibfnamefont {G.}~\bibnamefont {Weimann}}, \ and\ \bibinfo {author}
  {\bibfnamefont {W.}~\bibnamefont {Schlapp}},\ }\href {\doibase
  10.1103/PhysRevB.40.7874} {\bibfield  {journal} {\bibinfo  {journal} {Phys.
  Rev. B}\ }\textbf {\bibinfo {volume} {40}},\ \bibinfo {pages} {7874}
  (\bibinfo {year} {1989})}\BibitemShut {NoStop}%
\bibitem [{\citenamefont {Willett}\ \emph {et~al.}(1990)\citenamefont
  {Willett}, \citenamefont {Paalanen}, \citenamefont {Ruel}, \citenamefont
  {West}, \citenamefont {Pfeiffer},\ and\ \citenamefont
  {Bishop}}]{willett1990}%
  \BibitemOpen
  \bibfield  {author} {\bibinfo {author} {\bibfnamefont {R.~L.}\ \bibnamefont
  {Willett}}, \bibinfo {author} {\bibfnamefont {M.~A.}\ \bibnamefont
  {Paalanen}}, \bibinfo {author} {\bibfnamefont {R.~R.}\ \bibnamefont {Ruel}},
  \bibinfo {author} {\bibfnamefont {K.~W.}\ \bibnamefont {West}}, \bibinfo
  {author} {\bibfnamefont {L.~N.}\ \bibnamefont {Pfeiffer}}, \ and\ \bibinfo
  {author} {\bibfnamefont {D.~J.}\ \bibnamefont {Bishop}},\ }\href {\doibase
  10.1103/PhysRevLett.65.112} {\bibfield  {journal} {\bibinfo  {journal} {Phys.
  Rev. Lett.}\ }\textbf {\bibinfo {volume} {65}},\ \bibinfo {pages} {112}
  (\bibinfo {year} {1990})}\BibitemShut {NoStop}%
\bibitem [{\citenamefont {Willett}\ \emph
  {et~al.}(1993{\natexlab{a}})\citenamefont {Willett}, \citenamefont {Ruel},
  \citenamefont {Paalanen}, \citenamefont {West},\ and\ \citenamefont
  {Pfeiffer}}]{willett_enhanced_1993}%
  \BibitemOpen
  \bibfield  {author} {\bibinfo {author} {\bibfnamefont {R.~L.}\ \bibnamefont
  {Willett}}, \bibinfo {author} {\bibfnamefont {R.~R.}\ \bibnamefont {Ruel}},
  \bibinfo {author} {\bibfnamefont {M.~A.}\ \bibnamefont {Paalanen}}, \bibinfo
  {author} {\bibfnamefont {K.~W.}\ \bibnamefont {West}}, \ and\ \bibinfo
  {author} {\bibfnamefont {L.~N.}\ \bibnamefont {Pfeiffer}},\ }\href {\doibase
  10.1103/PhysRevB.47.7344} {\bibfield  {journal} {\bibinfo  {journal} {Phys.
  Rev. B}\ }\textbf {\bibinfo {volume} {47}},\ \bibinfo {pages} {7344}
  (\bibinfo {year} {1993}{\natexlab{a}})}\BibitemShut {NoStop}%
\bibitem [{\citenamefont {Ridley}(1988)}]{ridley1988}%
  \BibitemOpen
  \bibfield  {author} {\bibinfo {author} {\bibfnamefont {B.~K.}\ \bibnamefont
  {Ridley}},\ }\href {\doibase 10.1088/0268-1242/3/6/005} {\bibfield  {journal}
  {\bibinfo  {journal} {Semicond. Sci. Technol.}\ }\textbf {\bibinfo {volume}
  {3}},\ \bibinfo {pages} {542} (\bibinfo {year} {1988})}\BibitemShut {NoStop}%
\bibitem [{\citenamefont {Willett}\ \emph {et~al.}(1995)\citenamefont
  {Willett}, \citenamefont {West},\ and\ \citenamefont
  {Pfeiffer}}]{willett1995}%
  \BibitemOpen
  \bibfield  {author} {\bibinfo {author} {\bibfnamefont {R.~L.}\ \bibnamefont
  {Willett}}, \bibinfo {author} {\bibfnamefont {K.~W.}\ \bibnamefont {West}}, \
  and\ \bibinfo {author} {\bibfnamefont {L.~N.}\ \bibnamefont {Pfeiffer}},\
  }\href {\doibase 10.1103/PhysRevLett.75.2988} {\bibfield  {journal} {\bibinfo
   {journal} {Phys. Rev. Lett.}\ }\textbf {\bibinfo {volume} {75}},\ \bibinfo
  {pages} {2988} (\bibinfo {year} {1995})}\BibitemShut {NoStop}%
\bibitem [{\citenamefont {Willett}\ \emph
  {et~al.}(1993{\natexlab{b}})\citenamefont {Willett}, \citenamefont {Ruel},
  \citenamefont {West},\ and\ \citenamefont
  {Pfeiffer}}]{willett_experimental_1993}%
  \BibitemOpen
  \bibfield  {author} {\bibinfo {author} {\bibfnamefont {R.~L.}\ \bibnamefont
  {Willett}}, \bibinfo {author} {\bibfnamefont {R.~R.}\ \bibnamefont {Ruel}},
  \bibinfo {author} {\bibfnamefont {K.~W.}\ \bibnamefont {West}}, \ and\
  \bibinfo {author} {\bibfnamefont {L.~N.}\ \bibnamefont {Pfeiffer}},\ }\href
  {\doibase 10.1103/PhysRevLett.71.3846} {\bibfield  {journal} {\bibinfo
  {journal} {Phys. Rev. Lett.}\ }\textbf {\bibinfo {volume} {71}},\ \bibinfo
  {pages} {3846} (\bibinfo {year} {1993}{\natexlab{b}})}\BibitemShut {NoStop}%
\bibitem [{\citenamefont {Simon}(1996)}]{simon1996a}%
  \BibitemOpen
  \bibfield  {author} {\bibinfo {author} {\bibfnamefont {S.~H.}\ \bibnamefont
  {Simon}},\ }\href {\doibase 10.1103/PhysRevB.54.13878} {\bibfield  {journal}
  {\bibinfo  {journal} {Phys. Rev. B}\ }\textbf {\bibinfo {volume} {54}},\
  \bibinfo {pages} {13878} (\bibinfo {year} {1996})}\BibitemShut {NoStop}%
\bibitem [{\citenamefont {Nozi{\`e}res}(1997)}]{nozieres1997}%
  \BibitemOpen
  \bibfield  {author} {\bibinfo {author} {\bibfnamefont {P.}~\bibnamefont
  {Nozi{\`e}res}},\ }\href@noop {} {{\selectlanguage {English}\emph {\bibinfo
  {title} {Theory of Interacting {{Fermi}} Systems}}}}\ (\bibinfo  {publisher}
  {{Addison-Wesley}},\ \bibinfo {year} {1997})\BibitemShut {NoStop}%
\bibitem [{Note1()}]{Note1}%
  \BibitemOpen
  \bibinfo {note} {The constant has an oscillatory dependence on $d$. Our own
  calculation shows the same qualitative behavior as that shown in Ref~\protect
  \citep {simon1996a} but with quantitative differences. It indicates that the
  constant has a maximal value $3.7\times 10^{-4}$ at $d=0$ and a second peak
  $3.1\times 10^{-4}$ at $qd\sim 2.5$.}\BibitemShut {Stop}%
\end{thebibliography}%

\end{document}